\documentstyle[12pt,aasms4]{article}

\begin{document}

\title{Compact Nuclei in Galaxies at Moderate Redshift:II. Their Nature
and Implications for the AGN Luminosity Function}
\author{Vicki L. Sarajedini}
\affil{Steward Observatory, University of Arizona, Tucson, AZ 85721}
\authoremail{vicki@as.arizona.edu}
\author{Richard F. Green}
\affil{NOAO\footnote{The National Optical Astronomy Observatories
are operated by the Association of Universities for Research in
Astronomy, Inc., under Cooperative Agreement with the National
Science Foundation.}, P.O. Box 26732, Tucson, AZ 85726}
\authoremail{green@noao.edu}
\author{Richard E. Griffiths and Kavan Ratnatunga}
\affil{Carnegie Mellon University, Department of Physics,
5000 Forbes Ave., Pittsburgh, PA 15213-3890}
\authoremail{griffith@astro.phys.cmu.edu, kavan@astro.phys.cmu.edu}

\begin{abstract}

This study explores the space
density and properties of active galaxies to z$\simeq$0.8.
We have investigated the frequency and nature of unresolved
nuclei in galaxies at moderate redshift as indicators of nuclear
activity such as Active Galactic Nuclei (AGN) or starbursts.
Candidates are selected by fitting imaged galaxies
with multi-component models using maximum likelihood estimate techniques
to determine the best model fit.  We select those
galaxies requiring an unresolved point-source component in the galaxy
nucleus, in addition to a disk and/or bulge component, to adequately model
the galaxy light.

We have searched 70 WFPC2 images primarily from the Medium Deep Survey
for galaxies containing compact nuclei.
In our survey of 1033 galaxies, the fraction
containing an unresolved nuclear component $\geq$5\% of the
total galaxy light is 9$\pm$1\% corrected for incompleteness.
In this second of two papers in this series, we discuss the
nature of the compact nuclei and their hosts. 

We present the upper limit luminosity
function (LF) for low-luminosity AGN (LLAGN) in two redshift bins to z=0.8.
Mild number density evolution is detected
of the form $\phi$$\propto$(1+z)$^{1.9}$ for nuclei at
-18$\lesssim$M$_B$$\lesssim$-14.
The LFs appear to flatten at M$_B$$\geq$-16 and this
flatness, combined with the increase in number density,
is inconsistent with pure luminosity evolution.
Based on the amount of density evolution observed for these objects, we
find that almost all present-day spiral galaxies could have hosted a LLAGN
at some point in their lives.
We also comment on the likely contribution of these compact nuclei to the 
soft X-ray background.
\end{abstract}


\section{Introduction}

To better understand the nature of any class of extragalactic object, an
accurate knowledge of the luminosity function (LF) over a wide range of 
absolute magnitudes and covering a range of redshifts is necessary.
For active galactic nuclei (AGN) this has been attempted using
quasars at the brighter, primarily high redshift end and
Seyfert galaxy nuclei, considered to be their intrinsically fainter 
counterparts, at the low luminosity, low redshift end (Cheng et al. 1985;
Huchra \& Burg 1992).  
Understanding how the faint end of the AGN LF evolves
is of particular importance for determining the frequency and 
total space density of these objects at earlier epochs.
Models for quasar evolution can be
much better constrained with accurate knowledge of the shape
at the faint end.  Koo (1986) explains how many of the models
are difficult to descriminate when only luminous quasars are included since
this end of the LF is a power-law with a slope
that is almost identical at all redshifts.

Furthermore, AGN are likely contributors to the X-ray background
and several studies have determined the contribution of luminous
quasars to the diffuse background (e.g. Schmidt \& Green 1986).
How the low luminosity AGNs (LLAGNs) contribute to
the diffuse X-ray background has been a question of interest for some
time (Elvis et al. 1984; Koratkar et al. 1995).
An accurate understanding of the behavior of the faint AGN LF and its
evolution would help us to
determine the global significance of their contribution to the
soft X-ray background.

How then can LLAGNs be observed in higher redshift 
galaxies? The use of spectroscopic techniques becomes quite difficult at
moderate redshifts.  Even locally, the observation of broad emission 
lines, indicative of Seyfert 1 activity, or emission line flux ratios 
indicating Seyfert 2 or LINER activity requires that the nucleus dominate 
the galaxy light (e.g. Huchra \& Burg 1992) or have adequate spatial resolution 
for subtraction of the galaxy light from the nucleus (Ho et al. 1997a). 
At moderate redshifts LLAGNs become virtually
impossible to observe from the ground becoming indistinguishable 
from central regions of enhanced star formation or finite central 
density cusps of spheroidal components. 

The Medium Deep Survey (MDS) (Griffiths et al. 1994) is an HST key project
yielding two to four parallel WFPC2 exposures per week, each containing
$\sim$300 galaxies down to V$\sim$ 23.5 mag.
This survey provides an ideal sample of distant field galaxies for which
morphology and galaxy light profiles can be studied at
sub-kiloparsec resolution.  Typical galaxy redshifts extend to z$\simeq$0.8. 
The set of Cycle 4, 5 and 6 images consists of $\sim$150 fields with 
both V(F606W) and I(F814W) exposures.

This database provides a unique opportunity to search for morphological
evidence of AGN or other compact nuclear activity such as starbursts.
The nuclear activity of Seyfert galaxies manifests itself morphologically
as an unresolved point source in the nucleus of the galaxy.
This is due to the fact that most of the emission is originating from a small
region typically a few parsecs in diameter.  For Seyfert 2 nuclei the
emission is probably originating over the broader narrow-line region but
is still highly concentrated at the nucleus having nuclear
FWHM$\lesssim$200 pc (Nelson et al. 1996).  This is in agreement with
the recent WFPC2 study of Seyferts by Malkan et al. (1998).  They were able 
to resolve the nuclei of almost all local Seyfert 2 galaxies whereas most
local Seyfert 1 nuclei remained unresolved.    
The physical size of an unresolved region in a WFPC2 image at z$\leq$0.8
easily encompasses that for both types of Seyfert
nuclei.  Other enhancements of this size in a galaxy profile would
include starburst regions and nuclear HII regions.  Nuclear starbursts
typically have sizes of a few hundred parsecs (e.g. Weedman et al. 1981
for NGC 7714) and nuclear HII regions may be even smaller.
These stellar enhancements in a galaxy light profile would appear unresolved
over the redshift range of interest (0.2$\lesssim$z$\lesssim$0.8) unless
H$_o$ is very high.
The typical late-type spiral bulge, however, has a diameter
of $\sim$1 kpc (Boroson 1981) which is resolved in the HST images.

Our study includes galaxies
in 70 WFPC2 fields which have been modeled to search for unresolved nuclei
likely to be AGN or compact regions of
star formation, i.e. starbursts.  The galaxy modeling is based on maximum
likelihood estimates used to extract quantitative morphological and
structural parameters of the faint galaxy images.  All galaxies to
I$\lesssim$21.5 in 64 MDS fields and 5 Groth survey strip fields
(E. Groth, P.I., Prop ID 5090)
as well as the Hubble Deep Field to I$\lesssim$23.5 (Williams et al.
1996) have been modeled
in this way to reveal an unresolved nuclear component when present.
Selection of AGN using this technique results in a
magnitude-limited sample which is
less biased towards galaxies dominated by the nucleus than are
spectroscopically selected samples.  In this way, we probe the intrinisically
fainter population of AGN and starbursts out to intermediate redshifts
for the first look at how this population of objects evolves.

In Sarajedini et al. (1998) (Hereafter referred to as Paper I), 
we discuss in detail the selection of fields and the
procedure for fitting the galaxy light profiles.  We also describe
the criterion used for selecting those galaxies containing compact nuclei 
along with various simulations to determine the uniqueness of the 
galaxy model.  In addition, we present in Paper I the ground-based 
spectroscopic data for about one third of our sample and describe 
the procedure of estimating redshifts for the remaining sample with
$\sigma$$_z$$\leq$0.1 using V-I color, I magnitude, and the bulge-to-total
luminosity ratio.

In this paper we focus on the analysis of the sample
selected in Paper I.  We discuss the properties of the host galaxies
such as color, magnitude, size and Hubble type.  These properties
are then compared with those of local Seyferts and starburst galaxies. 
The colors of the nuclei are used to differentiate between Seyfert-like
nuclei and young starburst nuclei based on colors from representative
spectra of these objects.  Using this information,
we calculate the upper limit luminosity function
for LLAGN in the magnitude range -20$\leq$M$_B$$\leq$-14 in two redshift
bins out to z=0.8 and compare it with the LFs of local Seyferts and
moderate redshift QSOs.  We also comment on the likely contribution of
these nuclei to the soft X-ray background.

\section{Properties of the Host Galaxies}

The sample of 101 galaxies containing unresolved nuclei described in Paper I
is large enough to study, with some statistical significance,
the morphological and photometric characteristics of the nuclei
and host galaxies.  Here we examine the properites of the hosts
and compare them with local AGN and starburst host galaxies.  The
colors of the nuclei themselves are discussed in section 3 and are used
to answer questions concerning the nature of this population of objects.

\subsection{Host Galaxy Types}

When discussing the host galaxy types for the point source nuclei
galaxies in our sample, it is very important to determine how well the bulge
component is measured in these galaxies.  This measurement effect
was examined using
Monte-Carlo simulations described in Paper I.
The galaxies used in the simulation were originally fit with disk+bulge
models before the simulated nuclear point source was added.  Based
on this $a$ $priori$ disk+bulge fitting, we have a known
``input" bulge-to-bulge+disk (B/B+D) luminosity ratio which can be compared with
the measured B/B+D luminosity ratio when the simulated point
source is detected in the galaxy.

These simulations show that for detections of point sources
greater than 20\% of the total galaxy light, the bulge is never
detected regardless of the input bulge value.
Our fitting technique is unable to separate and measure the bulge and
point source components when the point source is a large contributor
($\geq$ 20\%) of the total galaxy light.
This is not the case for fainter point sources ($\lesssim$ 20\%) where
the bulge is more accurately measured.  We note
that in all of the cases where the input bulge-to-total was zero,
it was also measured as zero.

For point sources fainter than 20\% of the galaxy light where the
input bulge-to-total ratio is non-zero, we determine how often is it measured
accurately and how often is it measured as zero.  Figure 1a
shows the fraction of galaxies where no bulge is detected
for the simulated galaxies as a function of the actual input bulge-to-total
luminosity ratio.  The error
bars represent the Poisson statistics based on the number of points in
each bin.  From this figure we see that as the bulge becomes more
significant, it is more likely to be measured accurately and not
go undetected.

We can apply this statistical result
to the real sample of compact nuclei galaxies to determine what fraction
of these galaxies may actually contain a significant bulge where none was
measured in the fitted model.
Of the 101 compact nuclei galaxies in our sample, only 8 have point
source-to-total luminosity percentages greater than 20\%.  For these
8, no bulge is detected as we would expect based on the
simulations.  These galaxies, however, could contain a significant bulge
component which is not being properly measured due to the bright point
source nucleus.  A hidden bulge component in these
galaxies may be as much as 50\% of the total light but is probably not
dominating the host galaxy since spiral structure is seen upon examination
of the galaxy images.

The remaining 93 galaxies have point source-to-total luminosity
percentages less than 20\%.  Of these, 57 are measured to have
no bulge component.  We can estimate the fraction of the 57 that
are statistically likely to contain a bulge based on Figure 1a.
In this figure, the fraction of galaxies having measured bulge-to-bulge+disk
luminosity ratios between 0 and 0.2 is high, indicating
that a large fraction of host galaxies having bulge components
as large as 20\% will be measured as having no bulge.
Our 57 hosts with no detected bulge may contain bulges that are up to
20\% of the total host galaxy light.  Therefore, our first
grouping of host galaxy types are those with 0$\leq$(B/B+D)$<$0.2 since
it is difficult to differentiate these galaxy types with higher resolution.
There are a total of 79 hosts in this grouping. 

There are 9 galaxies measured with 0.2$\leq$(B/B+D)$<$0.5.  According
to Figure 1a, 59\% $\pm$9\% of galaxies with this size bulge will be
measured with no bulge when a point source (of less
than 20\% the galaxy light) is detected.  If 9 galaxies have
0.2$\leq$(B/B+D)$<$0.5 then as many as 9 to 19 galaxies of this
type will be measured with no bulge component.
Three host galaxies fall in the 0.5$\leq$(B/B+D)$<$0.8 bin.  Figure
1a indicates that 21\% $\pm$9\% of these galaxy types will be measured
as having no bulge.  Based on this percentage, we would expect one
additional host galaxy with 0.5$\leq$(B/B+D)$<$0.8 to be measured as
having no bulge component.

According to this analysis, the 57 host galaxies measured with
no bulge component are likely to be comprised of 9 to 19 galaxies with
0.2$\leq$(B/B+D)$<$0.5 and 1 galaxy with 0.5$\leq$(B/B+D)$<$0.8.
Removing these 10 to 20 galaxies from the B/B+D=0 bin allows us to
statistically redistribute the host galaxy types based on our previous
simulations.  The solid line in Figure 1b is the measured
B/B+D distribution for
the 101 host galaxies containing compact nuclei.  The dashed lines are the
statistically redistributed histograms where the 2 lines represent
the two extremes of the distribution.  Here we have
assumed that the 8 host galaxies with point source nuclei greater than
20\% of the galaxy light have bulges less than 50\% of the host galaxy
light and could lie entirely in the 0$\leq$(B/B+D)$<$0.2 bin
or the 0.2$\leq$(B/B+D)$<$0.5 bin.

The dashed lines in Figure 1b indicate that the
actual number of host galaxies having 0$\leq$(B/B+D)$<$0.2 is likely
to be from 58\% to 76\% of the 101 host galaxies in our sample.
Eighteen to 36\% may have 0.2$\leq$(B/B+D)$<$0.5.  Only 6\% are likely to
have B/B+D$\geq$0.5.  We note, however, that the incompleteness
estimates for detecting point sources in galaxies containing a substantial
bulge is high.  In Paper I we show that the completeness
is typically 25 to 40\%
for galaxies with 0.5$\leq$(B/B+D)$<$1.0.  Therefore, there are likely
to be 2.5 to 4 times as many point source nuclei present than were actually
found in galaxies with large bulges.

Figure 1b shows that even at the extremes,
the majority of host galaxies (58-76\%) have small bulges indicating
late-type spiral galaxy hosts.  One-fifth to one-third may be
early-type spirals with bulges greater than 20\% of the galaxy light but
less than 50\%.
We detect only 6\% with bulges greater than 50\% of the host galaxy
light.  Incompleteness corrections could increase this from 6\% to $\sim$10\%
(see discussion in the following section).

To compare our sample with local host galaxy types of Seyferts and starbursts,
we will assume a galaxy with B/B+D$\leq$0.2 is
a late type spiral (Sc to Sb) and 0.2$\leq$(B/B+D)$<$0.5 is
an early type spiral (Sa to S0).  Although Hubble type does not depend on
Bulge/Disk measurement alone we can estimate the approximate type
based on a study of the luminosity
distribution in nearby spiral galaxies (Boroson 1981).
The CfA Seyfert galaxies studied in McLeod \& Rieke (1995) had Hubble types
ranging from S0 to Sc with no elliptical hosts.
45\% were found in late type spirals and the remaining 55\% in
early type spirals.  These Seyferts, although considered low-luminosity
AGN, are likely to have intrinsically brighter nuclei than our sample
since they were selected spectroscopically.  However, a similar distribution
is found in
the Ho et al. (1997a) study of ``dwarf" Seyfert nuclei in nearby
galaxies.  They find about 8\% of the dwarf nuclei reside in ellipticals,
while 46\% are in S0 to Sa galaxies and another 46\% reside in Sb through 
Sm galaxies.  In contrast, Balzano (1983) finds that for starburst nuclei 
in local galaxies, 3\% are in ellipticals,
30\% are in early type spirals, and 67\% are in late type spirals.
Locally, starburst galaxies appear to favor late type spirals more so
than dwarf Seyfert nuclei and this causes the local starburst host type
distribution to more closely resemble our host galaxy type distribution.
However, our distribution would also be consistent with some combination
of local Seyfert nuclei hosts and starburst nuclei hosts which is the
expected mix of object types in our sample. 

As part of this study, we also note the galaxies in our survey for which
we obtained spectra and compare their Bulge/Total measurements
with Hubble-types as estimated from their spectra.  Kennicutt (1992)
presents integrated spectra of nearby galaxies and compares emission
and absorption features with Hubble class for normal and peculiar
galaxies for a range of galaxy types.  He finds that early-type
galaxies from E's to Sa's have similar absorption feature
spectra.  H$\alpha$ + [NII] and [SII] become prominent for Sb's and later types.
In Sc's through irregulars, we see many more emission lines such as
[OIII], H$\beta$ and [OII].

In our spectra, 10 galaxies had wavelength coverage for H$\alpha$
detection.  H$\alpha$ emission was observed in 9 of those galaxies all
of which had Bulge/Total values less than 0.06, consistent with
late-type spirals.  The one without H$\alpha$ detected had
Bulge/Total=0.3 which is consistent with an early-type spiral galaxy.
For these cases, our Bulge/Total measurements appear to be consistent
with spectroscopic classifications for these galaxies.

Out of the 29 spectra for which redshifts could be determined, 23
showed emission lines while 6 were only absorption spectra.  Ninety-one
percent of the emission line spectra galaxies had Bulge/Total$\leq$0.2 and
67\% of the absorption spectra had Bulge/Total$\geq$0.2.  Again,
these statistics indicate that the majority of galaxies in our sample
with measured spectra have spectral types consistent
with their Bulge/Total measurements.

\subsection{Comparison with Other Survey Galaxies}

\subsubsection{Morphology}

It is interesting to compare the host galaxies of our compact nuclei
with the entire sample of 1033 galaxies from the MDS, Groth strip,
and HDF.  As mentioned before, the measurement of Bulge/Total luminosity
ratio gives us an estimate of the probable Hubble type for a galaxy.
Figure 2 is the histogram of Bulge/Total values for the entire sample
of galaxies in our survey.  The hatched region is the true normalized histogram
while the dotted line represents the fraction of galaxies found in
each Bulge/Total bin defined in the previous subsection.
This dotted line can be compared to the host
galaxy Bulge/Bulge+Disk distributions in Figure 1b.
All of the possible host galaxy distributions in Figure 1b
are weighted towards later-type galaxies as compared to
the total galaxy sample.  In Figure 2, a larger fraction of host galaxies have
0$\leq$(B/B+D)$<$0.5 than survey
galaxies.  Although the largest fraction of survey galaxies also
have 0$\leq$(B/B+D)$<$0.5, there are many more with
0.5$\leq$(B/B+D)$<$1.0 than what is seen in the host galaxy sample.

This apparent discrepancy could
be attributed in part to incompleteness in detecting nuclei in galaxies
with large bulges.  We can correct for this effect using
conservative completeness
estimates calculated in Paper I (90\% completeness for
(B/B+D)=0, 55\% completeness for 0$<$(B/B+D)$\leq$0.4, 43\% for
0.4$<$(B/B+D)$\leq$0.8, and 38\% for (B/B+D)$>$0.8).  Adjusting our
numbers in each B/B+D bin accordingly indicates an increase
in the number of host galaxies with 0.5$\leq$(B/B+D)$<$1.0 from 6\%
to $\sim$10\% $\pm$2.5\%.  This fraction is still significantly 
lower than the $\sim$15\% of
all survey galaxies of this type.  Based on this distribution, the host
galaxies containing compact nuclei do appear to favor galaxies with
bulges contributing less than 50\% of the total galaxy light.

\subsubsection{Size and Magnitude}

We investigate how the compact nuclei galaxies vary in
angular size and apparent magnitude from the entire sample of
survey galaxies in Figure 3.
Figure 3a shows the angular size (natural log of the
half-light radius)
in arcsec vs. the apparent I band magnitude of the
survey galaxies (open circles) and the host galaxies (filled circles).
The typical limiting magnitude for MDS and Groth strip fields at
I$\simeq$21.5 is apparent in the diagram.
Most of the fainter galaxies
are from the Hubble Deep Field.
The smallest host galaxies are also from the HDF and are
as small as 2 pixels (0.2$\arcsec$) in radius.
It appears that few of the large but faint galaxies, possibly
low surface brightness galaxies, appear to host point source nuclei.
This trend is further revealed in Figure 3b which is the normalized
distribution of the half-light radii for the host galaxies (hatched region)
and the non-host galaxies (solid line).  The host galaxy distribution
peaks at a smaller half light radius (hlr$\simeq$0.45$\arcsec$)
than the non-host galaxies (hlr$\simeq$0.72$\arcsec$) and also
cuts off at a half-light radius of $\sim$1.9$\arcsec$.
A KS test conducted on both the size and magnitude distributions
of the two populations indicates that they are drawn from different
parent populations at the 80\% significance level.
A possible explanation is that
these galaxies may not contain enough mass to
sustain a starburst nucleus or AGN.  Recent work by Schombert (1998)
has shown that HI-rich, low surface brightness galaxies have a greater
chance of hosting LLAGN than other magnitude limited, late-type galaxies.  
This seems to indicate that the gas mass is an important factor
in the existance of an active nucleus.
Additionally, some mechanism is
usually required to transport fuel to the nucleus.  Mihos et al. (1997)
shows through
numerical simulations that low surface brightness galaxies are stable
against the growth of bar instabilities, inhibiting a strong inflow of
gas to the galaxy center for fueling a nuclear starburst or AGN.
Although our nuclei are intrinsically faint, this observation implies
that some limiting mass or ability to transport matter to the nucleus is
necessary for their formation and duration.

\subsubsection{Axis Ratios}

Optically selected AGN have been observed to avoid
edge-on spirals (Keel 1980; McKleod \& Rieke 1995).  Figure 4
shows the distribution of axis ratios for the disks of
our host galaxies.  We have included only those
hosts that are disk-dominated, containing bulges less than 20\%
of the total galaxy light.  The distribution is normalized by
dividing the axis ratio distribution of
host galaxies by the same distribution for the total sample of
survey galaxies which are also disk-dominated.
This normalizes the axis ratio distribution so that each bin represents
the fraction of all disk-dominated galaxies with the specified axis
ratio which contain compact nuclei.
There is an observed drop in the fraction of hosts
with disk axis ratios (b/a) $\lesssim$0.4.
However, this drop is much less significant than that observed
in the spectroscopically selected CfA Seyferts.
Lawrence and Elvis (1982) found that an inclination
bias does not appear in hard X-ray selected samples of AGN.  They
conclude that selection biases seen in other samples are due to
obscuration in a flattened disk parallel to the plane of the
host galaxy.  Although the axis ratio distribution of our data reveals a
slight bias against edge-on host galaxies, our morphological selection
technique appears to be less affected by this type of obscuration.

\subsubsection{Galaxy Color}

Finally, we examine how the colors of the host galaxies
compare with the non-host galaxies.  First we must understand the
accuracy of the host colors.  Our Monte Carlo simulations in Paper I 
show that the bulge component of a host galaxy is often undetected when
a nuclear point source is fit.  At the same time, these simulations show
that the point source magnitudes are well determined within the
errors.  Therefore, the undetected bulge component does not appear to
be mistakenly measured as part of the point source.  How then does the missing
bulge component effect the magnitude and consequently the color measurement
for the host galaxy?  We again look at the Monte-Carlo simulations from
Paper I to compare the $input$ galaxy colors to those measured for
the galaxy when a simulated point source is detected in the nucleus.
For 389 galaxies in the simulation where the point source nucleus was
detected (in both or only one filter) the 
distribution of color errors was centered around zero with
$\sigma$$_{(V-I)}$=0.37.
This indicates that we are not seeing a systematic offset between
the measured and true color
of the host galaxy which might be expected if the bulge were being
completely ignored in the model fitting.  Instead, when the bulge
is not detected, it seems to
be absorbed in the disk component such that the host galaxy color is
not systematically affected.  Keeping in mind the typical errors in
determining the host galaxy color, we compare them to those of galaxies
in our survey without unresolved nuclear components.

Figure 5a is the color-magnitude diagram for the galaxies in our survey
where the open circles are non-host galaxies and the filled circles
are the host galaxies.  We have removed the color contribution of
the nucleus from the integrated galaxy color to produce the host galaxy
V-I color.
In order to directly compare the sample of host galaxies with a similar
representation of the survey galaxies,
a random subset of the 1033 survey galaxies was selected based on the
host galaxy distribution in Figure 1b (short-dashed line).
This allows us to compare our host galaxy colors to those of
survey galaxies morphologically similar to our hosts.
Figure 5b shows the normalized
histogram of the V-I colors for the compact nuclei host galaxies
(hatched region) where the subset of similar survey
galaxies is represented by the solid line.
A KS test for these color distributions yields a probability of 0.43,
where low KS probabilities indicate that two populations are not
drawn from the same parent distribution.  This result
indicates that the two distributions are not statistically different.
The mean color for the host galaxies is $\mu$$_{(V-I)}$=0.97
while the mean survey galaxy color is $\mu$$_{(V-I)}$=1.03.
In general, we find that the host galaxy colors are like those
of other morphologically similar galaxies in the field
with a slight tendency to be bluer.

\subsection{Host Galaxy Absolute Magnitudes}

Using the spectroscopic and photometric redshifts for each galaxy
containing an unresolved point source nucleus, we can examine the
absolute magnitudes of the host galaxies as they compare with other
galaxies.  Figure 6a is the histogram of absolute rest-frame
B magnitudes
for the 101 host galaxies which contain unresolved nuclei using H$_o$=75
km/s/Mpc and q$_o$=0.  The B-band luminosities for the
galaxies were derived from their V magnitude, V-I color and redshift.
This information
is used to select a non-evolving model spectral energy distribution from
the set described in Gronwall and Koo (1995).  The models are based
on those of Bruzual and Charlot (1993).  The k-corrections are small
(typically $\lesssim$0.3 mag) since the V magnitude is roughly rest-frame
B at the median redshift of our survey (z$\simeq$0.35).

Our luminosity distribution peaks
at M$_B$$\simeq$-20.4 which is approximately L$^*$.
For comparison with the following samples, this corresponds to a peak at
M$_B$$\simeq$-21.3 with H$_o$=50 km/s/Mpc.  The luminosity distribution
of normal spiral galaxies peaks at M$_B$=-20.6 (H$_o$=50 km/s/Mpc)
(Christensen 1975), somewhat fainter than our sample of compact nuclei
galaxies.  Interestingly, Yee (1983) found the peak luminosity of
Seyfert hosts from the Markarian survey to be M$_B$=-21.3 $\pm$0.8 which
is strikingly similar to our sample.  From this observation he
argues that the Seyfert phenomenon tends to occur more often in the
luminous end of the spiral population.  According to the present study,
this trend may also hold true for galaxies hosting compact nuclei.
This brighter absolute magnitude for AGN hosts is also seen, though to a
lesser degree, for X-ray selected AGN where the peak luminosity is
M$_B$=-20.9 (Kotilainen \& Ward 1994).

We can compare our distribution to the dwarf seyfert nuclei hosts
of Ho et al. (1997a) (Figure 6b) and the distribution of galaxies containing
HII nuclei from  Ho et al. (1997b) (Figure 6c) which were derived
using H$_o$=75 km/s/Mpc.  These samples provide
an interesting comparison since the $nuclear$ luminosities are closer
to those of our sample than other spectroscopically selected AGN.
The luminosity distributions
are similar among the three samples.  Although the peak
luminosities are close, our sample does appear to drop off faster on
the low luminosity side of the distribution.  This is most likely
caused by the nature of our magnitude-limited survey as it is observed
over a range in redshift.
As we observe at higher redshifts, intrinsically fainter galaxies are missed.
Accounting for this discrepancy, we observe no evidence for evolution
in the luminosity of AGN or starburst host galaxies out to z$\simeq$0.8.

\section{The Colors of the Point Source Nuclei: Clues to Their Nature}

Model fitting of the point source component allows us to separate the
color of the nucleus from that of the host galaxy.  Of the 101
nuclear point sources detected, 93 were detected in the I filter image,
80 were detected in the V filter image, and 72 were detected in both
filters allowing for a V-I color to be determined.  This result leaves 29
point sources with no color information.  There are a few reasons for
not detecting the point source in one of the filters such as 1) the point
source is too faint in one filter, 2) the point source does not
converge properly in one filter making its magnitude unreliable,
and 3) the image is too noisy in one filter for a point source
to be detected.  Of these 29, 5 were not detected in the other filter
because of non-convergence of the model, 7 were not detected because
of additional noise elements in that filter, and 17 were not
detected because the point source was too faint in that filter.
For these 17, we can determine a limit for the point source
color assuming the point source in the other filter is fainter than
1\% of the total galaxy light.  For the other 12 point sources detected
only in one filter, we cannot constrain the nuclear color.

Figure 7a shows the V-I colors of the point source nuclei vs.
their host galaxy V-I colors.  Here we have plotted the 72 points having
color information with their appropriate error bars.  The 17 points with
a point source color limit, due to non-detection in one filter, are
marked with an arrow.
Within the errors of the point source colors, the majority of point source
nuclei are the same color as the underlying host galaxy.
The large errorbars on the point source colors cause the correlation
coefficient between the point source and host galaxy colors to be
small indicating a low probability of correlation.
In Figure 7b we show the generalized histogram of color differences
between the
host and point source where point sources bluer than the host are
positive and those redder than the host are negative.  This histogram
is made by allowing each point to be represented as a gaussian with
the calculated 1$\sigma$ error defining the width of the gaussian.  The
histogram is the summation of these gaussians.
It does not include the 17 points with only color limits.
The peak of this distribution occurs at $\mu$=-0.05, essentially zero,
and the FWHM of the distribution is 1.52.
The nuclear colors are in general the same as the host galaxy colors
with a large spread in the distribution.
Nuclei which appear bluer than the host can
be explained as either young starburst nuclei or unobscured AGN.
However, a nucleus can appear redder than the host if there is
sufficient dust present or excess star formation in the disk causes
the underlying galaxy to appear bluer.

It is also interesting to see how the nuclear colors compare with the
color of the bulge where it is detected in both the
V and I filters.  Again, we first must understand the errors in the
bulge colors as determined from the Monte-Carlo simulations.
Based on comparison of the input bulge colors of the simulation galaxies
to the measured bulge colors when a point source nucleus is detected,
we determine the error in bulge color to be $\sigma$$_{V-I}$=0.85.
This error is large but not unexpected since the bulge component appears
to be the most difficult to measure accurately when a point source
nucleus is detected in the galaxy.  Again, there is no systematic
color difference in the bulges observed in the simulation galaxies when a point
source is detected.  We apply the determined color error for bulges to
our data to investigate any trends in the relationship between the
nuclear and bulge colors.

Figure 8a is the point source color vs. the bulge color
for the 17 galaxies where both a point source and bulge color were measurable.
We have applied the determined bulge color error and point source color
errors.  Because these errorbars are so large, this figure is not useful
for studying individual cases.  In Figure 8b we show the generalized histogram
of the color difference between the point source and bulge components.
The histogram is made in the way described for Figure 7b.  This figure
allows us to observe any trends in the color difference while taking
into account the errors in the individual points.
The mean color difference between the bulge and the point source
nucleus is (V-I)$_{bulge}$--(V-I)$_{nucleus}$$\simeq$0.5.   This
indicates that in general,
the nuclear colors are bluer than the bulge.  This is expected
if the bulge represents an older population than the
nucleus or if the nucleus has typical non-thermal source colors.

Studying the nuclear colors allows us to look more deeply into
the question of their nature.  In the case of
detection in a bulge dominated system, the color of the nucleus plays
a role.  As discussed above, we require the nucleus to be
a different color from that of the underlying bulge to ensure that it is not
merely an unresolved region of the bulge.
Are there objects, besides Seyferts and starbursts, that could
appear unresolved in an elliptical or bulge dominated galaxy and
would be composed of a different stellar population from that of the bulge
so that it would appear a different color?  One additional possibility
is a merger remnant.  Theoretical and observational evidence suggests
that major mergers, where the galaxies are of almost equal mass,
form elliptical remnants (Hibbard et al. 1994 and references
therein).  In the case of the merger remnant NGC 3921, an unresolved
nucleus is observed in an r$^{1/4}$ profile galaxy (Schweizer 1996).
The color gradient of the remnant indicates a redder nucleus than the
remaining galaxy body.
Such remnant nuclei could exist in our sample, although the small number (6\%)
of elliptical hosts suggests they would be a minor contributor to the
population.  There is also the possibility that minor merger remnants
in the centers of galaxies could reveal themselves as unresolved nuclei
of a different stellar population than the host galaxy.  It is
unclear at this time how often such an object might occur in our survey.

With the spectroscopic and photometric redshifts of each host determined,
we can examine any trends in the nuclear colors with redshift.  This
information is important in determining the true nature of the nuclei.
Figure 9 shows the V-I colors of the nuclei plotted against their
determined redshift.  Errorbars in the redshift direction clearly
differentiate between those objects with photometric redshifts
($\sigma$$_z$$\simeq$0.9) and those with spectroscopic redshifts.
Open circles are those point sources with only limiting color information.
To interpret this figure, we must understand how the colors of
starbursts, AGN or any other stellar population thought to be present
in our nuclear sample behave as a function of redshift.  This modeling is
explored in the following section.

\section{Synthetic V-I colors for AGN and Star Clusters: Comparison
with the Nuclear Colors}

The V-I colors of the point source nuclei in our sample provide important
information to help us to understand their nature.
To determine if the nuclei are starbursts or AGN, we must first understand
the colors of these different objects.  Most starburst and AGN
photometry (other than that of bright QSO's) has been done at low redshifts.
To simulate the effects of
K-corrections on the object colors and therefore provide a more
realistic comparison with our data, we use representative spectra to calculate
synthetic V-I colors for AGN and starbursts with the HST F606W (V) and 
F814W (I) filters.

To simulate colors for the AGN nuclei, we use spectra from Kalinkov et
al. (1993) for Seyfert 1, 1.5 and 2 galaxies
where the nuclear activity dominates the spectrum.  These spectra, which
were produced by averaging the spectra of several reliably classified AGN in
each class, are shown in Figures 10a, 10b and 10c.
Spectra of star clusters of various ages are used to simulate the 
appearance of certain isochronic stellar populations present in the nuclei
sample.  The spectra are from E. Bica and are available as part of a database
of spectra for galaxy evolution models (Leitherer et al. 1996).
Figure 10d is an integrated spectrum of HII regions, representing a
very young population or current starbursting nucleus.  Figure 10e
represents a star cluster with age$\simeq$10 Myr and
[Z/Z$_{\sun}$]$\simeq$-0.25.  Figure 10f represents a star cluster
with age$\simeq$25 Myr and [Z/Z$_{\sun}$]$\simeq$-0.4.  The ages of the
star cluster spectra shown in Figures 10g through 10i are 80 Myr, 200-500
Myr and 1-2 Gyr, respectively.  Their respective metallicities are
-0.5, -0.6 and -0.5.   
 
To simulate the effects of redshift in our sample, the
spectra were stepped incrementally in redshift to z$\sim$0.8 and
the IRAF task CALCPHOT was used to determine the V-I colors
of each object at a range of redshifts.  CALCPHOT utilizes the
appropriate HST filter response which is convolved with the
spectra to produce accurate synthetic colors.
Figure 11a shows the V-I colors for the 3 types of Seyfert
galaxies as a function of redshift.  The solid line is the Seyfert 2
color, the short dashed line is the Seyfert 1.5 and the long dashed
line is the Seyfert 1.  The majority of nuclei are consistent
with the Seyfert colors within the errorbars. 
The two spectroscopically identified Seyfert 1s in our sample
at z=0.45, V-I=0.89 and z=0.99, V-I=0.22 are consistent
with the Seyfert 1 colors.   Of the 89 point source
nuclei with measurable colors or color limits, 8 are bluer than the Seyfert
colors by more than 1 $\sigma$ while 8 are redder.  The latter sample, however,
could be explained with moderate amounts of reddening due to dust.
 
Figure 11b shows the V-I colors for the various
star cluster spectra shown in Figures 10d through 10i as a function of redshift.
The solid line is the integrated HII region color.
This locus represents the color
a young starburst would have in our sample.  The color varies with
redshift as different emission lines in the spectrum pass through
the V and I filters.  The blueward shift in the HII region
color at z$\simeq$0.35 is caused by the 4959$\AA$ and
5007$\AA$ [OIII] and H$\beta$ emission lines moving into the V filter.
The remaining lines are specified in the figure caption and
represent the other star clusters with increasing age corresponding to
redder colors.  The majority of the
point source nuclei are consistent with the colors of intermediate
aged star clusters.  Five of the 8 galaxies which were too blue
to be consistent with Seyferts fall within 1 $\sigma$ of the HII region color
at z$\lesssim$0.4.
Another one of these 8 point sources at z=0.60, V-I=-0.10 has a color
consistent with that of a 25 Myr old cluster.  Together, these 6 point
sources have colors more consistent with those of star clusters than
Seyfert nuclei.  The other 2 point sources too blue to be Seyferts are
also inconsistent with the star cluster colors.

The synthetic colors show that the vast majority of the nuclei in
our sample have colors consistent with Seyfert-like nuclei {\it and}
intermediate aged star clusters.  A small number
(6 out of 89) have colors too blue to be Seyferts but are consistent
with young or intermediate aged star cluster colors.  These 6 are
likely to be starburst nuclei.  The main population,
however, cannot be identified explicitly as either Seyfert-like
or starburst-like based on the V-I color alone.   We assume that
the remaining sample of 95 point sources, which includes the 12 point sources
with no measurable color, consists of some combination of starbursts
and Seyfert nuclei.  The remaining 95 point sources
represent an upper limit sample of AGN-like
nuclei.  In the next section, this sample is used to determine an
upper limit on the AGN luminosity function at low luminosities.

\section{The Number Density and Luminosity Function for Unresolved Nuclei}
 
The luminosity function of AGN as a function of redshift provides much needed
information about how this population of objects changes with time.
Understanding the population as a whole may allow us to
understand better the physics of QSOs.  In addition, accurate knowledge
of the AGN LF is important
for understanding their contribution to the soft X-ray background.

The bright end of the QSO LF has been well studied as a function of
redshift (e.g. Hartwick \& Schade 1990,
and references therein).  To study the faint end, Seyfert galaxies
have been observed locally through spectroscopic surveys (Cheng et al. 1985;
Huchra \& Burg 1992).
However, these surveys tend to include only bright Seyfert nuclei
that dominate the galaxy light.  Fainter nuclei would
not be spectroscopically selected since the host galaxy light would
dilute the nuclear emission.  Morphologically, such faint nuclei can be
detected if the nuclear light can be disentangled from the host galaxy.
In this study we have separated the nuclear and galaxy light
using 2-dimensional maximum likelihood galaxy modeling.
This method allows us to study unresolved nuclear sources in galaxies out
to moderate redshifts to fainter limiting magnitudes than previously
observed.
 
Before constructing the luminosity function, we first compare
the fraction of galaxies containing compact nuclei
with the fraction of Seyfert galaxies determined locally
and at moderate redshifts.
As mentioned previously, Huchra \& Burg (1992) (hereafter HB)
have studied local Seyferts from the
CfA redshift survey and find the percentage of galaxies containing Seyfert
1 and 2 nuclei to be $\sim$2\%.  Their study included Seyfert nuclei
extending to M$_B$$\lesssim$-17.  Maiolino \& Rieke (1995)
examined a sample of Seyfert galaxies closer than the
CfA sample and were able to detect fainter nuclei since the
nuclear spectra were less diluted by the host galaxy light.
They found the percentage of Seyferts to be 5\% and possibly as high
as 16\%.  The majority of the additional objects in this study
were Seyfert 2s which are underluminous in the optical with respect
to type 1 Seyfert nuclei.  Active nuclei were detected in this study to
M$_B$$\lesssim$-16.
In Paper I, we find the fraction of all
galaxies in our survey containing compact nuclei to be 16$\pm$3\%
including nuclei $\geq$3\% of the galaxy light
corrected for incompleteness.  This fraction is very similar to that
found by Maiolino \& Rieke for local Seyferts, although our sample includes
nuclei at least a magnitude fainter.
 
Ho et al. (1997a) find the
percentage of local galaxies containing faint Seyfert nuclei
to be 11\% with LINERs (low-ionization nuclear emission line regions)
occupying an additional 19\% of galaxies.
They find the fraction of galaxies containing HII nuclei
to be 42\% of all galaxies.  These nuclei are extremely faint based on
their emission line luminosities (L(H$\alpha$)$\leq$10$^{40}$ ergs s$^{-1}$).
Many of these nuclei are too faint for morphological detection as
a point source in the host galaxy and were identified
based on emission lines alone.  These
results are intriguing as they suggest that the possible fraction
of local galaxies containing low-luminosity active nuclei may be as
high as 30\% (including LINERS).  Since optical integrated magnitudes
for these AGN and HII nuclei have not been determined, we cannot directly
compare the fraction of galaxies in the present study containing compact
nuclei with the fraction of galaxies containing faint AGN and HII nuclei
from Ho et al.
 
At moderate redshifts (z$\leq$0.3), Tresse et al. (1996) have searched
for evidence of nuclear activity in the Canada-France Redshift
Survey (CFRS) and find that 17\% of all galaxies
have emission line flux ratios consistent with active nuclei galaxies.
Since these objects are spectroscopically selected, this survey
contains only those nuclei which dominate the host galaxy light
such that their emission lines are not greatly diluted by the underlying
galaxy.  When they correct for possible stellar absorption of the Balmer
lines due to the host galaxy, the fraction of galaxies displaying activity
is 8\%. Their results are somewhat
inconsistant with those of Maiolino \& Rieke since the fraction of
higher redshift galaxies containing active nuclei identified by Tresse
et al. appears to be less than that for local Seyfert nuclei.
Such inconsistencies highlight
the need for space density studies of LLAGN at
moderate redshift.
 
\section{The Upper Limit Luminosity Function for LLAGN}
 
In section 4 we selected a subsample of our compact nuclei which
have colors consistent with those of Seyfert nucleus-dominated
galaxies based on synthetic photometry of representative spectra.
However, the colors are also consistent with intermediate aged
starburst nuclei making this subsample an upper limit estimate
on the number density of AGN in this survey.  This subsample can
be used to construct the upper limit luminosity function for
faint AGN out to redshifts of z$\simeq$0.8, providing the
first look at the shape and parameters of the AGN LF in this luminosity
and redshift regime.

\subsection{LF Calculation}
 
To calculate the luminosity function for our sample of 95 compact
nuclei with colors consistent with AGN-like nuclei, we use the 1/V$_A$
technique described fully in Schmidt and Green (1983) where V$_A$ is
the accessible volume in which each object can be observed.
This technique allows us to calculate the space density of the
compact nuclei as a function of their absolute magnitudes and
redshifts.  This quantity is symbolically defined as $\phi$(M,z).
To calculate absolute magnitudes from the nuclear apparent magnitudes
we use the equation
\begin{equation}
M_B=B-5logA(z)+2.5(1+\alpha)log(1+z)-5log(c/H_o\times10^6)+5
\end{equation}
where B is the apparent magnitude, c is the speed of light in
km/s, H$_o$ is the Hubble constant in km s$^{-1}$ Mpc$^{-1}$, and
$\alpha$ is the spectral index defined by
\begin{equation}
F_{\nu}\propto\nu^{\alpha}
\end{equation}
The quantity A(z) is the bolometric luminosity distance defined as
\begin{equation}
A(z)=z\left\{ 1+{z(1-q_o)}\over{[(1+2q_oz)^{1/2}+1+q_oz]}\right\}
\end{equation}
The term 2.5(1+$\alpha$)$log$(1+z) in equation 1 represents the effect
of the redshift on measurements through a fixed color band (i.e. the
K-correction).  Since we do not have apparent B magnitudes, we
determine them from the apparent V magnitude and the spectral energy
index, $\alpha$, according to the following equation
\begin{equation}
B-V = {\alpha}2.5log_{10}(\lambda_B/\lambda_V)
\end{equation}
where $\lambda$$_B$=4500$\AA$ and $\lambda$$_V$=6060$\AA$ for the HST
filters.  We assume the width of the filter bands to be comparable.
With $\alpha$=-1.0, this equation yields B-V=0.32.
We use $\alpha$=-1.0 or -0.5 in later calculations for direct
comparison with other LFs in the literature.
Based on the spectra for Seyfert-nucleus
dominated galaxies described in section 4, their rest frame B-V colors are
0.34, 0.50 and 0.71 for Seyfert 1, 1.5 and 2 galaxies respectively.
Our choice of $\alpha$ for the sake of comparison may be
somewhat blue for the typical object in our survey, however, the choice
of magnitude bin sizes is much larger than this error in the magnitude.
 
The absolute magnitude of each point source in the sample is determined
according to the above equations.  With the absolute magnitude we
determine the maximum redshift at which the point source would be observable
in each of our 70 WFPC2 fields.
The limiting point source magnitude per field is determined for each object
by calculating its apparent magnitude had it been detected in a galaxy at the
limiting galaxy magnitude of each field in the study.  This point source
limiting apparent magnitude per field is based on the observed
percentage of galaxy light contained within the nucleus.
The maximum redshift per field is then used, in combination with
the total area of the field in steradians, to determine the comoving
volume over which each object is observable in each survey field.
The summation of the volumes for each field is the accessible volume
for that particular object.  The quantity $\phi$ is then
determined as a function of
absolute magnitude and redshift through the following summation
\begin{equation}
\phi(M,z)dM=\sum_{j=1}^{n} {1\over{V_a^j}}
\end{equation}
for z$_j$$\in$(z$_1$,z$_2$) and M$_j$$\in$(M$_1$,M$_2$)
where M$_1$ = M - dM/2, M$_2$ = M + dM/2, z$_1$ = z - dz/2,
and z$_2$ = z + dz/2.  The integer $j$ denotes individual galaxies
in each absolute magnitude and redshift bin.
In this way, the number density is determined for a discrete luminosity
interval and a discrete redshift interval.
 
Because of the large errorbars in redshift for the photometrically
determined objects in our sample, an additional consideration is
made in determining the luminosity function.  Each individual
object is considered as a Gaussian distribution in redshift space.
In this way, each object is spread over its appropriate
luminosity range in the LF calculation.  The determination of
$\phi$ is then computed by
\begin{equation}
\phi(M,z)dM= \sum_{k=1}^{n} {(Weight)\over{V_a^k}}
\end{equation}
so that only the appropriately weighted portion of each point is
considered in the number density calculation in each specific luminosity
and redshift interval.  The value of $k$ represents the individual fractions
of each galaxy so that $n$ is the number of galaxies multiplied
by the number of divisions in redshift space over which each
object is distributed.

We also correct for incompleteness in our LF based
on our analysis from Paper I.  For each
point source in the survey, the level of incompleteness is determined
depending on the point source-to-total luminosity ratio of the object.
This factor is then multiplied with 1/V$_a$ for
each object in the
survey so that the final LF reflects the incompleteness correction.

Although we attempt to correct for incompleteness in our survey,
there are certain classes of hosts we may miss 
in a systematic fashion due to limitations in galaxy modeling
and our selection technique for detecting compact nuclei. 
Such classes could include faint dwarf ellipticals with steep profiles.
However, we note that for galaxies near L$^*$ the sample
should be reliably determined by total magnitude.

\subsection{Comparison with the Local Luminosity Function for Seyferts}
 
Figure 12 is the luminosity function for nuclei having colors
consistent with Seyferts.  Here we have plotted log($\phi$) in units
of \# Mpc$^{-3}$  mag$^{-1}$ vs. the absolute B magnitude.
The apparent V magnitudes were used to determine B magnitudes
using equation 4.  Table 1
provides the data used to construct this figure,
including the number of data points
in each bin $n$ and the number corrected for incompleteness
$n$$_{corrected}$.
The luminosity function shown is for point sources contributing
$\geq$3\% of the total galaxy light.  Figure 12
is the luminosity function with H$_o$=100 km s$^{-1}$ Mpc$^{-1}$,
q$_o$=0.5 and $\alpha$=-1.0.  The thin solid line represents the
LF at 0$<$z$\leq$0.4 and the dashed line represents the LF
at 0.4$<$z$\leq$0.8.
The mean redshift of the low z bin is $<$z$>$=0.31 and for the
high z bin is $<$z$>$=0.57.
We compare our LF with that of HB for Seyfert
1 and 2 galaxies from the CfA redshift survey (thick solid line).

Our LFs appear to be in good agreement with that of HB and maintain
the same shape within the overlapping magnitude region, 
-18.5$\geq$M$_B$$\geq$-20.5.
Recent X-ray observations support the HB number density.
The {\it Einstein} Extended Medium
Sensitivity Survey (EMSS) provides an X-ray selected sample of LLAGN
for which the optical luminosity function has been derived
(Della Ceca et al. 1996).  They used 226 broad-line AGN from this
survey with measured redshifts and V magnitudes.  Their LF extends
to the same limiting absolute magnitude as that of HB and
they find good agreement with their number density for broad-line AGNs.
Here we have compared the present data with the HB LF for both Seyfert 1 and
2 type galaxies.  Although the EMSS result is only for Seyfert 1 galaxies,
it confirms the number density of broad line LLAGNs locally through a
different selection method supporting the HB number density at faint absolute
magnitudes.

To provide a more careful check on the possibility of increase or
decrease between our LFs and that of HB within their overlapping luminosity
range, we consider the number density of nuclei in our sample which are
similar to the Seyfert nuclei of HB.  Granato et al. (1993) have shown
that Seyfert nuclei from the CfA survey and other Markarian Seyferts
have nuclei which contribute 20\% to 100\% of the total galaxy light.
If we limit our LF to include only our nuclei that comprise more than
20\% of the galaxy light, and are therefore more like those objects included
in the HB study, we find the number density to decrease by a factor
of 1.25 from that shown in Figure 12 for the LF at M$_B$=-19 and -20.
This amount of decrease is still consistent with our number density being
the same as that of HB in these luminosity bins.  We can further
attempt to improve the comparison by correcting for possible incompletion
in the HB data due to the observed axis ratio bias of their sample illustrated
in Maiolino \& Rieke (1995).  Accordingly, the HB data may be incomplete by
a factor of 2.  Correcting their LF for this level of incompleteness
and comparing with our corrected LF, the number densities at M$_B$=-19 and -20
remain the same within the errors and are consistent with no change in the
number density of LLAGN over this luminosity range.  A statistical test
is done by computing the density weighted 1/V$_a$'
number density (discussed in the following paragraph) for
our nuclei in the overlapping luminosity bins and determining if the
amount of density increase is statistically signficant when compared to
the HB number density in the same luminosity range.
This test reveals that the two datasets are consistent
with no density evolution within the errors.
 
Focusing now on the faint end of our LFs we notice
that our upper limit luminosity
functions for LLAGNs appear to flatten in both the high
and low redshift bins at M$_B$$\geq$-16.  The implication
of this observed flattening with regard to quasar evolution models
is discussed in detail later in this chapter.
There does not appear to be evidence for a significant increase in number
density between our high redshift LF and the low redshift LF within
the Poisson errors.  However, the high redshift LF consistently lies above
the low redshift LF.  This is most obvious at M$_B$$\geq$-18.
To determine if the density increase is statistically
significant, we use the $<$V'/V$_a$'$>$ method where V$_a$' is the density
weighted accessible volume (Schmidt \& Green 1983).  Since our LFs are
nearly parallel, we assume luminosity independent density evolution
of the form $\rho$(z)=(1+z)$^{\beta}$.  The mean value of V'/V$_a$' is
\begin{equation}
<V'/V_a'>=0.5\pm(12n)^{-1/2}
\end{equation}
for objects distributed according to the assumed density equation.
We begin our test by assuming no density evolution ($\beta$=0)
which yields $<$V'/V$_a$'$>$=0.602.
We then increase $\beta$ until $<$V'/V$_a$'$>$=0.5$\pm$0.043.  
The value of $<$V'/V$_a$'$>$ falls within this range ($\leq$0.543)
at $\beta$=1.9, which represents
the least amount of statistically significant density evolution
to describe our data at z$\leq$0.8.
 
Because our data represent upper limits of the AGN LF, this apparent
increase in number density at the faint end should be carefully interpreted.
If our selection technique systematically includes more objects at
high redshift than at low redshift, the increase may not be real.
However, we have shown that the nuclear region unresolved by
HST increases slowly with redshift, varying only
by 250 parsecs in diameter over the range 0.2$\lesssim$z$\lesssim$0.8.
One would need a significant population of nuclear objects having
actual sizes between 300 and 550 pc in diameter so that they would
appear unresolved in the higher redshift bin but appear
resolved and therefore not included in the low redshift bin.
 
A test can be performed to determine if the increased number density observed
for our unresolved nuclei could be due to a population of nuclear objects which
appear unresolved at high z while remaining resolved at low z.  Based on the
distribution of bulge sizes in our study, we simply count the number of
bulges that would appear unresolved if the number density of bulges
in our survey remained constant down to bulge diameters of zero.
In this way, we assume an upper limit number density for the small bulge
population.  We compare the number of bulges which would be
observed in the lower redshift half of the survey as compared to
the upper redshift half where we divide equally the volume of space.
Equal volumes of space are measured at 0$\leq$z$\leq$0.447 and
0.447$\leq$z$\leq$0.8.  In this experiment, we find that the number of
unresolved bulges would remain constant within the Poisson errors
between the two redshift bins of equal volume.
We find that the number of nuclear point sources
in our survey increases by a factor of 1.65 between the low redshift
and the high redshift bins which is a statistically signficant number
density increase within the Poisson errors.
This simple test indicates that the observed increase
in number density could not be due to unresolved bulge-like objects
in our survey.

Another test of the number density increase is conducted by performing
the $<$V'/V$_a$'$>$ calculation over a smaller range in redshift.  By
limiting the redshift range, we reduce any effect on the observed number
density evolution caused by marginally resolved nuclear objects
becoming unresolved and adding to the number density at higher z.  If
we limit ourselves to performing this calculation for objects observed
out to z=0.4, we find that the least amount of density evolution
necessary to describe the observations is consistent with that required
to describe the data out to z=0.8.
 
For the reasons we have outlined here, we interpret the increase
in number density as resulting from an actual increase in the population
of LLAGNS and nuclear starburst regions at -18$\leq$M$_B$$\leq$-14.
This observation poses the following question:
if we observe an increase in number density at the faint end of our LF
(M$_B$$\leq$-18) but no increase at M$_B$=-19 to -20 as compared with
HB, what are the implications for the global evolution of active nuclei?
One explanation is that the population mix at the faint end is different
from that at the bright end.  Where we overlap with the LF of HB, our nuclei
are likely to be more like the traditional Seyfert 1 galaxies.  This is
supported by the spectroscopic identifications of our two brightest nuclei.
However, as we look at less luminous nuclei, we are probing a mix of
fainter Seyfert 2s and LINERS.  This population
reveals an increase in number density with redshift out to z$\simeq$0.8
whereas the brighter Seyfert 1 nuclei do not show any number density
evolution out to this redshift.  In the following section, we discuss
further the implications of this scenario for comparing our observations
with the intrinsically brighter QSO LFs.
 
\subsection{Comparison with QSO Luminosity Functions}
 
Figure 13 is the luminosity function with H$_o$=50 km s$^{-1}$ Mpc$^{-1}$,
q$_o$=0.5 and $\alpha$=-0.5.  Again, the solid line represents the
LF from 0$<$z$\leq$0.4 and the dashed line is for 0.4$<$z$\leq$0.8.
Here we compare with the LF for QSO's from the compilation by
Hartwick and Schade (1990) for QSO's having  0.16$<$z$\leq$0.4 (solid line
at higher luminosities) and 0.4$<$z$\leq$0.7 (dashed line at higher
luminosities).  These redshift bins are roughly comparable to ours.
Table 2 provides the data used to construct these LFs.

This figure is useful for studying how the LF at fainter absolute magnitudes
compares with the bright end and allows us to investigate the nature of
the luminosity function evolution.
There is some indication in this figure that the shape of the
LF changes between the low and high redshift bins when we consider both the
low (this study) and high (Hartwick \& Schade) luminosity points.
If we assume that our upper limit LFs represent LLAGNs which are of the same
nature as the brighter QSOs,
any change in the overall shape of the LF would argue against both
pure density evolution and pure luminosity evolution for these objects.
Pure density evolution of the quasar LF has been ruled out by
results from a large number of quasar surveys (e.g. Cheney \& Rowan-Robinson
1981).  In pure density evolution, the
past LF is the same in shape as the local LF but shifted to higher
densities.  An overview of quasar evolution models
can be found in Koo (1986).  In pure luminosity evolution,
a uniform shift to brighter luminosities is observed while the LF shape
is preserved.  Under these assumptions, the combined LFs in Figure 13
do not appear to favor
either of these scenarios although we examine this question further below.

There have been many analytical representations for the shape of the
quasar LF proposed in the literature.  The double power-law model
of Boyle et al. (1988) is well-constrained and adequately describes the
data at z$\lesssim$2.2.  This model can be written as
\begin{equation}
{\phi(L,z)} = {{\phi_*/L_z}\over{(L/L_z)^{\beta_l}+(L/L_z)^{\beta_h}}}
\end{equation}
where the redshift dependence, L$_z$, is
\begin{equation}
L_z = L_*(1+z)^{-(1+\alpha)}exp[-(z-z_*)^2/2\sigma_*^2]
\end{equation}
as in Pei (1995) where pure-luminosity evolution is assumed and
$\alpha$ is the spectral index.
In equation 8 $\beta$$_l$ is the power-law index for the faint
end while $\beta$$_h$ represents the bright end.  To better compare
our LF with the brighter quasar LF from Hartwick \& Schade, we
extrapolate this model, fitted to their data, to lower luminosities.

Figure 14a shows the LFs from Figure 13 with the double power-law
model fit to the high luminosity data shown as the dotted line.
The fitted parameters from Pei (1995) are $\beta$$_l$=1.64$\pm$0.18,
$\beta$$_h$=3.52$\pm$0.11, z$_*$=2.75$\pm$0.05, $\sigma$$_*$=0.93$\pm$0.03,
log(L$_*$/L$_{\sun}$)=13.03$\pm$0.10, and
log($\phi$$_*$/Mpc$^{3}$)=-6.05$\pm$0.15.
These parameters are the best fit to the Harwick \& Schade LFs at various
redshift bins.  Here we have calculated the model at the median redshift
for each of the two redshift bins to overlay with the LFs.  Although the
fit to the high luminsity LFs is very good, this model does not fit our LFs
at lower luminosities.  The model predicts lower number counts than our
observations in the low redshift bin at -19$\leq$M$_B$$\leq$-16
and in the high redshift bin at -19$\leq$M$_B$$\leq$-17.  We also
note that the model predicts lower number counts than that observed by HB
as inferred from the agreement in Figure 12 between their
LF and our data over the common absolute magnitude range.
 
Because our luminosity function is only an upper limit on the
AGN LF at moderate redshifts, it is not necessarily in disagreement
with the extrapolated number counts of the model fits.
As described previously, even if we assume that the nuclei in
our survey are non-thermal in nature,
our LF is likely to include an increased population of low-luminosity AGNs
such as LINERS and faint Seyfert 2 nuclei.
The extrapolated results of this model using the fitted parameters described
above may be consistent with the population of faint QSOs and Seyfert 1
galaxies but may underestimate the total number of AGN including
Seyfert 2s and LINERS at M$_B$$\geq$-22.
 
In Figure 14b we show the same LFs with the double power-law model
fitted to both sets of data using $\chi$$^2$ minimization.  The
fitted parameters are $\beta$$_l$=1.6,
$\beta$$_h$=3.4, log(L$_*$/L$_{\sun}$)=12.5 and
log($\phi$$_*$/Mpc$^{3}$)=-5.2 holding z$_*$ and $\sigma$ constant
at the fitted values of Pei.  The flattening of the LF in the faint
data is not well fit by this model.  The $\chi$$^2$ of the fit to this
combined data set indicates a worse fit than that obtained for
the bright quasar data alone in several redshift bins.
We notice that the largest change in the fitted parameters occurs
in the normalization and inflection point of the double power-law
model and not in the slopes.  Our two brightest LF
bins, which are consistent with the HB local LF, largely influence the
fit in this manner.
If the two data sets being combined here are actually representing
the same type of active nucleus, this result indicates that the
double power-law
model assuming pure luminosity evolution is not the
best representation of the data.
The transistion region between the low and high luminosity data appears to
require a model with additional variables for adequate fitting of this portion
of the LF.
 
This result leads to the following question:
without considering the bright quasar data from Harwick \& Schade,
how well does pure luminosity evolution (PLE) model our faint LF over the
redshift range to z=0.8?  We fit a simple power-law to our low and
high redshift LFs and determine the amount of PLE necessary to adequately
model the data.
 
A power-law is fit to both the high and low redshift LFs from
-18$\lesssim$M$_B$$\lesssim$-14 of the form
\begin{equation}
log\phi(z) = const+{\gamma}M(z)
\end{equation}
where $\gamma$ is the slope of the log$\phi$-M relation and
is found to be 0.079$\pm$0.066.  We find the mean increase in
log$\phi$ over this magnitude range is 0.4.  This $\Delta$log$\phi$
divided by $\gamma$ gives the change in magnitude,
$\Delta$M=5.1$^{+25.7}_{-2.3}$.  For pure luminosity evolution
\begin{equation}
{\Delta}M = 2.5{\delta}log((1+z_{high})/(1+z_{low}))
\end{equation}
Here, $\delta$ is the amount of PLE necessary to account for the
observed density increase in a power-law parameterization in (1+z).
Using the mean z for each of our LF bins, we determine
$\delta$=25.8$^{+130.1}_{-11.6}$.  This simple calculation
shows that the amount of PLE necessary to account for the observed density
increase is unreasonably large and unlikely to explain the
observations.
 
The fact that our LF is only an upper limit probably does little to remedy
the flatness of the faint end of the AGN LF.  Any population of objects
which may be included in our survey, such as intermediate
aged starburst nuclei or minor merger remnants, is likely to have
a luminosity function with a negative slope at the faint end similar to that
observed in the normal galaxy LF.  If this is the case, subtracting an
LF of this shape from our upper limit would result in further flattening
of the faint end of the AGN LF.

\section{The Number of LLAGN in Local Spiral Galaxies}
 
In this section we address the implications the observed number density
of compact nuclei in our survey has for
the lifetimes of LLAGN and the fraction of present-day spiral galaxies
that may host AGN of low luminosities.  We first make the assumption that
the observed number density increase with redshift is caused by the
evolution of LLAGNs.
As we will describe further in the following section,
we assume a Gaussian density evolution equation for these objects consistent
with the observed mild density evolution to z=0.8 but turning over at z=2.3
(see equation 16).  At their peak density at z=2.3, the
total number of unresolved nuclei is 5.9$\times$10$^{-3}$ Mpc$^{-3}$ mag$^{-1}$
with H$_o$=100 km s$^{-1}$ Mpc$^{-1}$.
If our sample consists entirely of LLAGNs, what fraction of
present-day
galaxies are likely to have contained an LLAGN at some point in their
lives? Typically our nuclei are found in spiral galaxies with absolute
magnitudes in the range -20.5$\leq$M$_B$$\leq$-18.5.
Using the LF of Marzke et al. (1994) for spiral galaxies, the number
of local spirals in this absolute magnitude range is 9$\times$10$^{-3}$ per
Mpc$^3$.  This result implies that almost all present-day spiral
galaxies contain LLAGN with absolute magnitudes of 
M$_B$$\simeq$-16 at some epoch, some of which may now be dormant.
The total number density at z=1.0, which requires less extrapolation of our
observations, indicates that $\sim$40\% of present-day spiral galaxies
contain LLAGN in this magnitude range.
If we assume the observed luminosity of these objects is due to disk accretion
onto a black hole, we can determine if the amount of matter necessary to
fuel an AGN of this luminosity over a Hubble time is plausible.  We convert the
absolute magnitude to bolometric luminosity based on
L$_{bol}$/L$_B$ = 11.8 $\pm$4.3 for quasars from Elvis et al. (1994).
An order of magnitude approximation for the implied $\dot{M}$ is given
by
\begin{equation}
\dot{M} \sim L_{bol}(10^{-9} M_{\sun}/yr)/(10^{37} ergs/s)
\end{equation}
from Shapiro \& Teukolsky (1983).  For nuclei with M$_B$=-16,
$\dot{M}$ equals 8$\times$10$^{-4}$ M$_{\sun}$/yr.  After 15 Gyr,
the object accretes $\sim$10$^7$ M$_{\sun}$, comparable to the mass of
the black hole, which is a reasonable amount of matter for accretion.
This result has interesting implications on the search for AGN
in local galaxy nuclei.
 
\section{Implications For the X-Ray Source Surface Density}
 
Quasars have long been known as strong X-ray emitters and statistically
significant samples have been
selected through X-ray surveys (e.g. Giacconi et al. 1979).
How quasars and their low-luminosity counterparts contribute to
the diffuse X-ray background has been a question of interest for some
time.  In particular,
low-luminosity AGNs are considered good candidates for explaining
the origin of the soft X-ray background (0.2 - 4.0 keV).
 
Schmidt and Green (1986) derived the contribution of LLAGNs (M$_B$$>$-23)
to be 29\% of the observed background at 2 keV assuming no evolution in
the number density.  They note that any substantial evolution would
dramatically increase this value.  Other studies indicate that
30\% to 90\% of the background at 2 keV originates from AGNs
(Boyle et al 1993) but with the brightest AGN making up 30\%.
Could the rest of the soft background be comprised of LLAGN?
 
We calculate the likely contribution of the compact nuclei in our
survey to the X-ray
background based on assumptions about the X-ray nature of the nuclei.
Let us assume that the X-ray flux of our nuclei is most like that for
fainter AGN such as Seyfert 2s and LINERS.
From the {\it Einstein Observatory} Imaging Proportional Counter (IPC), soft
X-ray spectra for 22 Seyfert 2s and some LINERS in the range of 0.2 to 4.0 keV
have been obtained (Kruper et al. 1990) and the X-ray luminosity for
these objects is 10$^{42\pm0.72}$ erg s$^{-1}$.
In order for these galaxies to be spectroscopically classified as
Seyfert 2s or LINERS, the
nucleus must contribute a large fraction of the total galaxy light.
From the CfA redshift survey, we know that spectroscopically selected Seyferts
have nuclei contributing between 20\% and $\sim$100\% of the total galaxy light
(Granato et al. 1993).  Because our nuclei are roughly an order of magnitude
fainter, we assume their X-ray flux is also an order of magnitude
fainter and is likely to be closer to 1$\times$10$^{41\pm0.72}$ erg s$^{-1}$.
This rough estimate is in good agreement with the soft X-ray flux of 5 LLAGNs
measured with ROSAT {\it HRI} (Koratkar et al. 1995).  They range in
x-ray luminosity from 10$^{40.2}$ to 10$^{41.4}$ erg s$^{-1}$.
 
We first assume that the LFs we observe consist entirely of LLAGN and
that there is mild number density evolution of
these objects over the absolute magnitude range
-18$\lesssim$M$_B$$\lesssim$-14.  Since the LF is relatively parallel over
this magnitude range in both redshift bins, we allow the number
density to be independent of luminosity for these objects where
\begin{equation}
log\phi = 1.9log(z+1)-4.09
\end{equation}
for H$_o$=50 km s$^{-1}$ Mpc$^{-1}$, q$_o$=0.5 and $\alpha$=-0.5 and
where $\phi$ is in units of Mpc$^{-3}$ mag${-1}$.
 
To determine the total X-ray flux for these objects, we integrate
the number density $\phi$(z) multiplied by the X-ray flux out to z$_{max}$.
\begin{equation}
I_{Tot} = \int_{0}^{Z_{max}} \phi(z) \frac{L_x}{4 \pi A^{2}(z)} \frac{dV}{dz} dz
\end{equation}
where I$_{Tot}$ is the total X-ray flux from the point source nuclei.
The volume element is defined in Avni (1978) as
\begin{equation}
dV/dz = \omega(c/H_o)A^2(z)(1+z)^{-3}(1+2q_oz)^{-0.5}
\end{equation}
We assume a Gaussian form for the number density evolution consistent with
the mild evolution observed out to z=0.8 (equation 13) but
reaching a maximum near z=2.3, the apparent
peak in the quasar number density from Schmidt, Schneider \& Gunn (1991).
This is written as
\begin{equation}
\phi(z) = 1.03\times10^{-3}(1+z)^{-0.5}exp(-(z-2.3)^2/2)
\end{equation}
in units of Mpc$^{-3}$ mag${-1}$.
The integration out to z$_{max}$=4.0 over 4 magnitude bins
(-18$\leq$M$_B$$\leq$-14)
yields a total X-ray flux of 10$^{-8.84\pm0.72}$ ergs s$^{-1}$ cm$^{-2}$
sr$^{-1}$ for the 0.2 - 4.0 keV range.
 
A recent measurement of the X-ray background using ROSAT
(Chen, Fabian \& Gendreau 1997) found an intensity of
1.46x10$^{-8}$ ergs s$^{-1}$ cm$^{-2}$ sr$^{-1}$ in the 1 - 2 keV range.
Hasinger et al. (1993) find an X-ray background intensity of 1.25x10$^{-8}$
ergs s$^{-1}$ cm$^{-2}$ sr$^{-1}$ for this energy range.  To convert
between the measured energy ranges for comparison, we assume a power-law
X-ray spectral index of $\alpha$$_x$=-0.5 based on {\it Einstein Observatory}
IPC observations of low luminosity AGNs
consisting primarily of Seyfert 2s and LINERS (Kruper et al. 1990).
The conversion factor from the 0.2 - 4.0 keV to the 1 - 2 keV range
is 0.267.  Assuming an X-ray
background intensity of 1.36x10$^{-8}$ for 1 - 2 keV
(an average of the two measured
values), our nuclei between -18$\leq$M$_B$$\leq$-14 contribute
up to 15\% of the X-ray background.
If we assume a steeper spectral index of $\alpha$$_x$=-1.5, as
suggested by Koratkar et al. (1995) for LLAGNs, our nuclei contribute
up to 10\% of the X-ray background. 
 
%We also calculate the contribution of these nuclei to the X-ray
%background using an optical-to-X-ray flux conversion from ROSAT.
%Based on 283 AGN from the ROSAT All-Sky Survey (Bade et al. 1995),
%the median log(f$_x$/f$_B$) is 0.55 with a range of $\sim$$\pm$1.
%The ROSAT energy range is 0.1 - 2.4 keV.
%We convert our M$_B$ magnitude to X-ray luminosity using this ratio
%and sum the total X-ray luminosity per absolute magnitude bin
%over the range -18$\leq$M$_B$$\leq$-14.  This total L$_x$ is then
%used in the integral in equation 14 to determine the total
%X-ray intensity for the nuclei in our sample.  Using this conversion,
%we find the X-ray intensity of our nuclei to be 10$^{-7.91\pm1}$
%ergs s$^{-1}$ cm$^{-2}$ sr$^{-1}$.  This corresponds to a mean contribution
%of 30\% of the X-ray background intensity at 1 - 2 keV although the uncertainty
%in the optical-to-X-ray flux allows for the total X-ray intensity
%to exceed the soft X-ray background.
 
Because we do not have actual X-ray flux measurements for the galaxies
in our sample, these estimates of their contribution
to the soft X-ray background are based on assumptions about the
X-ray nature of these objects.  Although we cannot say with certainty
what their true contribution is, these results suggest that LLAGN
may contribute a relatively significant portion
of the soft X-ray background but would not be able to make up
the bulk of X-ray emission at these energy levels.
Further study of the X-ray properties
of faint AGN would provide tighter constraints on their likely contribution
to the diffuse X-ray background.

\section{Conclusions}
 
This study explores the space
density and properties of active galaxies to z$\simeq$0.8.
We have investigated the frequency and nature of unresolved
nuclei in galaxies at moderate redshift as indicators of nuclear
activity such as AGN or starbursts.  The
main results and conclusions are as follows: 

$\bullet$ Compact nuclei appear to favor host galaxies with bulges contributing
less than 50\% of the galaxy light.  The majority of hosts have small
bulges consistent with late-type
spiral galaxies.  The distribution of host galaxies more closely
resembles that for nearby starbursting galaxies than
local dwarf Seyfert nuclei but could be consistent with some
combination of these two populations.
 
$\bullet$ The host galaxies
of compact nuclei have smaller half-light radii as compared to
the entire sample of survey galaxies.  Very few of the
large, faint galaxies in the survey contain compact nuclei.  This may
be a result of low surface brightness galaxies being stable against
the growth of bar instabilities which may help fuel nuclear activity.
The host galaxy colors are generally like those
of other morphologically similar galaxies in the field
with a slight tendency to be bluer. 
We also note that compact nuclei avoid
host galaxies with low disk axis ratios ((b/a) $\lesssim$0.4) although
to a much lesser degree than spectroscopically selected samples of Seyferts.
This slight effect may be
caused by obscuration in a flattened disk parallel to the plane of the
host galaxy or by the main disk containing dust extending
into the galaxy center.
 
$\bullet$ The distribution of absolute magnitudes for the host galaxies
peaks at M$_B$$\simeq$-21.3 with H$_o$=50 km/s/Mpc
which is very similar to that for local Seyfert hosts.
This distribution is consistent with 
that for local LLAGN and HII nuclei hosts found by Ho et al. (1997a
and 1997b).

$\bullet$ 
%The V-I colors of the nuclei can be used to help determine their nature
%by comparing them with the colors of Seyfert nuclei and other
%stellar populations of various ages.  Using spectra from the literature,
%we calculated synthetic colors
%for typical Seyfert 1, 1.5 and 2 galaxies dominated by the nucleus as
%well as HII nuclei and intermediate aged stellar populations as
%proxies for possible starburst nuclei or minor merger remnants in
%our sample.  
Most of the nuclei have colors consistent with
Seyfert nuclei and also consistent with intermediate age star clusters.
A small number have colors too blue to be Seyferts but are consistent
with young or intermediate aged star clusters and are
likely to be starburst nuclei.
 
$\bullet$ Using the subsample of nuclei having colors consistent with
Seyfert and intermediate age starburst nuclei, we construct
the upper limit luminosity function for
faint AGN out to redshifts of z$\simeq$0.8 providing the
first look at the shape and parameters of the AGN LF in this luminosity
and redshift regime.  Our upper limit LF compares well with that
for Seyferts in the CfA redshift survey (Huchra \& Burg
1992) with no evidence for an increase in the number density
over their common magnitude range.  However,
a mild increase in the number density
of compact nuclei is detected between our low ($<$z$>$=0.31)
and high ($<$z$>$=0.57) redshift LFs
of the form $\phi$$\propto$(1+z)$^{1.9}$ for nuclei at
-18$\lesssim$M$_B$$\lesssim$-14 (H$_o$=50 km/s/Mpc).
The mild number density evolution at the faint end of our LFs could
be the result of an increase in the fraction of Seyfert 2 nuclei
and LINERS in this luminosity range.
 
$\bullet$ Both the low and high redshift LFs
for our data appear to flatten at M$_B$$\geq$-16.  When compared with
the bright quasar LFs from Harwick \& Schade (1990), the overall
shape of the LF appears to change from low to high redshift arguing
against pure density and pure luminosity evolution if our LFs represent
the intrinisically fainter counterparts of QSOs.
However, this apparent change in shape may be caused in part by the
possibility that our fainter LFs contain a greater fraction of Seyfert 2
nuclei and LINERS.
The flatness of our LF at these faint magnitudes and the increase in
number density is inconsistent with pure luminosity evolution within
our observations.  Additionally, we find
the number density of our nuclei combined with the observed mild
evolution indicates that almost
all present-day spiral galaxies could host LLAGN either active or dormant.
 
$\bullet$ Assuming the X-ray flux of our nuclei to be similar 
to that of Seyfert 2s and LINERs, we estimate
these nuclei in the range -18$\leq$M$_B$$\leq$-14 can contribute up to 
10 -- 15\%
of the soft X-ray background at 1 to 2 keV.
Further study of the X-ray properties
of faint AGN is needed to provide tighter constraints on their
likely contribution to the diffuse X-ray background.

\section{Future Work}
 
The results of this study provide the first look at the population
of LLAGNs at moderate redshifts.  Many additional questions can be
studied with this dataset to understand better the processes that
create and maintain nuclear activity in galaxies.
In the cases of both AGN and starbursts, mechanisms are required within the host
galaxy to provide fuel for the nuclear activity.  Many such mechanisms
have been proposed such as bars, galaxy-galaxy interactions, and other
morphological disturbances.  These types of disturbances have been
well studied in local groups of galaxies giving only the current
picture of how host galaxy morphology relates to nuclear activity.
Studying a population of active galaxies
at earlier epochs will help us understand how such fueling mechanisms
evolve on a global scale over several gigayears.
 
%The residuals from our fitting algorithm can be quantitatively
%examined for evidence of bars, rings and other disk asymmetries which indicate
%a nuclear fueling mechanism.
%Of particular interest is the presence of bars at earlier epochs in galaxies
%having nuclear activity.  The theoretical models of Friedli \& Benz (1993)
%predict that once a bar has channeled enough of the galaxy's mass into
%the central few hundred parsecs of the galaxy, it will dissolve on
%approximately gigayear timescales.  

For example, determining the frequency of bars in AGN/starburst hosts 
to z$\simeq$0.8 would allow us to trace the global evolution
of bars in initiating nuclear activity.  If a higher frequency of bars
is detected at high z while decreasing towards lower z, the typical
duration of a bar in an active galaxy as well as the likely formation
epoch can be estimated.
This scenario is consistent with the lack of a bar/AGN connection observed
locally (McLeod \& Rieke 1995; Ho et al. 1997c).
If the bar frequency remains constant with z, bars might be forming at a rate
close to the rate at which they dissolve.
Another globally interesting question can be answered by studying the
control sample of field galaxies.
Since many galaxies may have experienced episodes of rapid star
formation, tracing the evolution of bars and other asymmetries in the control
sample will help us understand when and how these perturbations have
affected the general population of galaxies.
We can also determine how the presence of a bar affects
the AGN/starburst luminosity.  
 
%Studying the incidence of close pairs and interactions among AGN/starburst
%host galaxies will also reveal information about how the fueling mechanisms
%of active nuclei evolve.  Gravitionally induced perturbations to 
%normal galaxies
%may provide the initiative for active nucleus fueling.  An increase in the
%number of hosts having nearby neighbor galaxies at moderate redshifts will
%indicate at what epoch gravitionally induced perturbations most efficiently
%initiated nuclear activity.  It will also be possible to study the
%relationship between the mass or luminosity of the neighbor galaxy
%and the luminosity of the active nucleus.
%The limiting magnitude for detecting galaxies in most WFPC2 parallel
%images is $\sim$2
%magnitudes below that for the fitted galaxies.  This allows us to
%search over a large range in magnitude for active galaxy neighbors
%so that faint as well as bright companions can be detected.

In order to more directly compare the results of this survey to
spectroscopic surveys,
the morphologies of these survey galaxies must be examined in the
same way as our HST imaged galaxies.  Such a study will bring to light the
differences and biases in these survey techniques.
High resolution ground-based images for a sample of local low-luminosity
Seyfert galaxies would allow us to study the nuclear regions and
determine the fraction
in which the AGN can be morphologically detected.  Likewise, HST
images are available for many fields where spectroscopic follow-up
has been obtained with the CFHT and Keck telescopes. 
Spectroscopic selection of active galaxies in these surveys can 
then be compared with the morphological evidence of an active nucleus.
The results of such studies will allow for
a more direct comparison of the number density of local AGNs to more
distant ones observed with HST.

%Acknowledgements

\acknowledgements

We would like to thank the referee and editor for comments and suggestions
which have improved the quality of this paper.  
The authors would like to thank John Huchra and 
Rogier Windhorst for helpful discussions and guidance on this project. 
We are also grateful to Caryl Gronwall
for providing software and models to estimate galaxy K-corrections.
In addition, we thank Yichen Pei for providing us with QSO luminosity
function models and model fitting software.  This work is based
on observations taken with the NASA/ESA {\it Hubble Space Telescope},
obtained at the Space Telescope Science Institute, operated by the
Association of Universities for Research in Astronomy, Inc.  This work
was supported in part by STScI grant GO-02684.06-87A to R. F. G. for
the Medium Deep Survey project.

%References

%Begin Figure Captions
\newpage

\centerline{FIGURE CAPTIONS}

\figcaption{a) The fraction of galaxies in the Monte-Carlo simulation where
the measured Bulge/Total is zero as a function of the input Bulge/Total
for the galaxy.  The errorbars are the Poisson statistics based on the
number of galaxies in each bin.  b)
The distribution of Bulge/Bulge+Disk measurements for
galaxies containing compact nuclei.  The solid line is the actual
number distribution.  The two dashed lines represent two extremes of
the distribution
after statistically correcting for bulge misclassification as described
in the text.}

\figcaption{Histogram of measured Bulge/Total values for the 1033 galaxies
in our survey (hatched region) normalized by 1:500.
The dotted line represents the total fraction
of galaxies in each Bulge/Total bin.}

\figcaption{a) The galaxy I magnitude vs. the natural log of the
half-light radius of the galaxy for all survey galaxies (open circles)
and those galaxies containing compact nuclei (filled circles).  b) The
normalized distribution of half-light radii for all survey galaxies
(solid line) and those containing compact nuclei (hatched region).}

\figcaption{Histogram of the axis ratios for galaxies containing compact
nuclei.  The histogram is normalized by dividing by the axis ratio
histogram for all spiral survey galaxies.}

\figcaption{a) The color-magnitude diagram for galaxies in our survey where
open circles represent non-host galaxies and filled cirles represent
those galaxies hosting a compact nucleus.  b) The normalized histogram
of galaxy V-I colors for the host galaxies (hatched region) compared with
the normalized histogram of galaxy colors for a representative group
of non-host galaxies of similar morphological type (solid line).}

\figcaption{a) Histogram of rest-frame B absolute magnitudes for the
host galaxies in our sample.  b) Absolute B magnitudes from
Ho et al. (1997a) of dwarf Seyfert nuclei hosts.  c) Absolute B magnitudes from
Ho et al. (1997b) of HII nuclei hosts.}

\figcaption{a) The point source V-I color vs. the host galaxies V-I color.
Points measured in only one filter are shown as color limits.
b) The generalized
histogram of the color difference between the host galaxy and the point
source.}

\figcaption{a) The point source V-I color vs. the bulge V-I color for
the 17 galaxies containing a measureable bulge and point source color.
b) The generalized histogram of the color difference between the bulge and
the point source nucleus.}

\figcaption{The redshift vs. the V-I color of the unresolved nuclei.
The open circles represent color limits for some nuclei.}

\figcaption{a) Seyfert 1 spectrum produced by averaging several
reliably classified Seyfert 1 galaxies (Kalinkov et al. 1993).
b) Seyfert 1.5 galaxy spectrum.  c) Seyfert 2 galaxy spectrum.
d) HII nucleus from E. Bica in Leitherer et al. (1996).
e) age$\simeq$10 Myr star cluster with [Z/Z$_{\sun}$]$\simeq$-0.25.
f) age$\simeq$25 Myr star cluster with [Z/Z$_{\sun}$]$\simeq$-0.4.
g) age$\simeq$80 Myr star cluster with [Z/Z$_{\sun}$]$\simeq$-0.5.
h) age$\simeq$200-500 Myr star cluster with [Z/Z$_{\sun}$]$\simeq$-0.6.
i) age$\simeq$1-2 Gyr star cluster with [Z/Z$_{\sun}$]$\simeq$-0.5.
The y-axes are in flux density units of ergs s$^{-1}$ cm$^{-2}$ $\AA$$^{-1}$
normalized at 5600$\AA$ to 10.}

\figcaption{a) V-I colors for the 3 types of Seyfert
galaxies as a function of redshift.  The solid line is the Seyfert 2
color, the short dashed line is the Seyfert 1.5 and the long dashed
line is the Seyfert 1.
b) V-I colors for various
star cluster spectra as a function of redshift.
The solid line is the integrated HII region color,
the short dashed line is the 10 Myr cluster,
the long dashed line is the 25 Myr cluster, the dot-short dashed line is the
80 Myr cluster, the dot-long dashed line is the 200-500 Myr cluster,
and the dotted line is the 1-2 Gyr cluster.}

\figcaption{The luminosity function of compact nuclei $\geq$3\% of
the total galaxy light.  The thin solid line represents the LF for nuclei
at 0$<$z$\leq$0.4 and the dashed line is the LF at 0.4$<$z$\leq$0.8.
The thick solid line is the LF for Seyfert galaxies from the
CfA redshift survey (Huchra \& Burg 1992).  Here we have
assumed H$_o$=100 km s$^{-1}$ Mpc$^{-1}$,
q$_o$=0.5 and $\alpha$=-1.0. }

\figcaption{The luminosity function of compact nuclei $\geq$3\% of
the total galaxy light.  The solid line represents nuclei
at 0$<$z$\leq$0.4 and the dashed line is for 0.4$<$z$\leq$0.8.
At higher luminosities we show the LF for QSO's from the compilation by
Hartwick and Schade (1990) for 0.16$<$z$\leq$0.4 (solid line at M$_B$$\leq$-22)
and 0.4$<$z$\leq$0.7 (dashed line at M$_B$$\leq$-22).
Here we have assumed H$_o$=50 km s$^{-1}$ Mpc$^{-1}$,
q$_o$=0.5 and $\alpha$=-0.5.}

\figcaption{a) The double power-law model
fitted to the bright quasar LFs (dotted line) using parameters
from Pei (1995).  b) The same model fitted to both the
bright LF and the faint LF (this data).  We assume
H$_o$=50 km s$^{-1}$ Mpc$^{-1}$, q$_o$=0.5 and $\alpha$=-0.5.}

%Begin Tables

\clearpage

\begin{deluxetable}{ccccccc}
\tablewidth{0pt}
\tablecaption{Luminosity Functions (H$_o$=100, q$_o$=0.5 and $\alpha$=-1.0)}
\tablehead{
\colhead{M$_B$} & \colhead{$\phi$} & \colhead{n} & \colhead{n$_{corrected}$} &
\colhead{$\phi$} & \colhead{n} & \colhead{n$_{corrected}$}}

\startdata
& \multicolumn{3}{c}{0 $<$ z $\leq$ 0.4} &
\multicolumn{3}{c}{0.4 $<$ z $\leq$ 0.8} \nl
\cline{2-4}
\cline{5-7}
-13.0 & -3.02 & 0.33 & 1.20 & -2.64 &  0.08 &  0.31 \nl
-14.0 & -2.65 & 2.26 & 6.13 & -2.26 &  0.63 &  2.44 \nl
-15.0 & -2.81 & 3.01 & 9.92 & -2.61 &  2.58 &  7.51 \nl
-16.0 & -2.94 & 3.17 & 9.10 & -2.58 &  8.79 & 21.43 \nl
-17.0 & -3.09 & 2.96 & 6.55 & -2.94 &  7.63 & 17.22 \nl
-18.0 & -3.64 & 0.99 & 1.90 & -3.35 &  3.50 &  5.46 \nl
-19.0 &       &      &      & -4.18 &  1.02 &  1.54 \nl
-20.0 &       &      &      & -4.21 &  1.06 &  1.61 \nl
\enddata
\end{deluxetable}

\clearpage

\begin{deluxetable}{ccccccc}
\tablewidth{0pt}
\tablecaption{Luminosity Functions (H$_o$=50, q$_o$=0.5 and $\alpha$=-0.5)}
\tablehead{
\colhead{M$_B$} & \colhead{$\phi$} & \colhead{n} & \colhead{n$_{corrected}$} &
\colhead{$\phi$} & \colhead{n} & \colhead{n$_{corrected}$}}

\startdata
& \multicolumn{3}{c}{0 $<$ z $\leq$ 0.4} &
\multicolumn{3}{c}{0.4 $<$ z $\leq$ 0.8} \nl
\cline{2-4}
\cline{5-7}
-13.0 & -3.81 & 0.06 &  0.21 &       &      &       \nl
-14.0 & -3.98 & 0.16 &  0.58 &       &      &       \nl
-15.0 & -3.87 & 0.64 &  2.27 & -3.10 & 0.50 &  1.94 \nl
-16.0 & -3.51 & 3.23 &  9.40 & -3.64 & 1.03 &  3.68 \nl
-17.0 & -3.73 & 3.47 & 11.12 & -3.38 & 7.55 & 19.76 \nl
-18.0 & -3.99 & 2.33 &  6.65 & -3.73 & 7.93 & 18.87 \nl
-19.0 & -4.11 & 2.97 &  5.02 & -4.07 & 5.24 &  8.69 \nl
-20.0 &       &      &       & -4.75 & 1.68 &  2.55 \nl
-21.0 &       &      &       & -5.02 & 1.34 &  2.04 \nl
\enddata
\end{deluxetable}

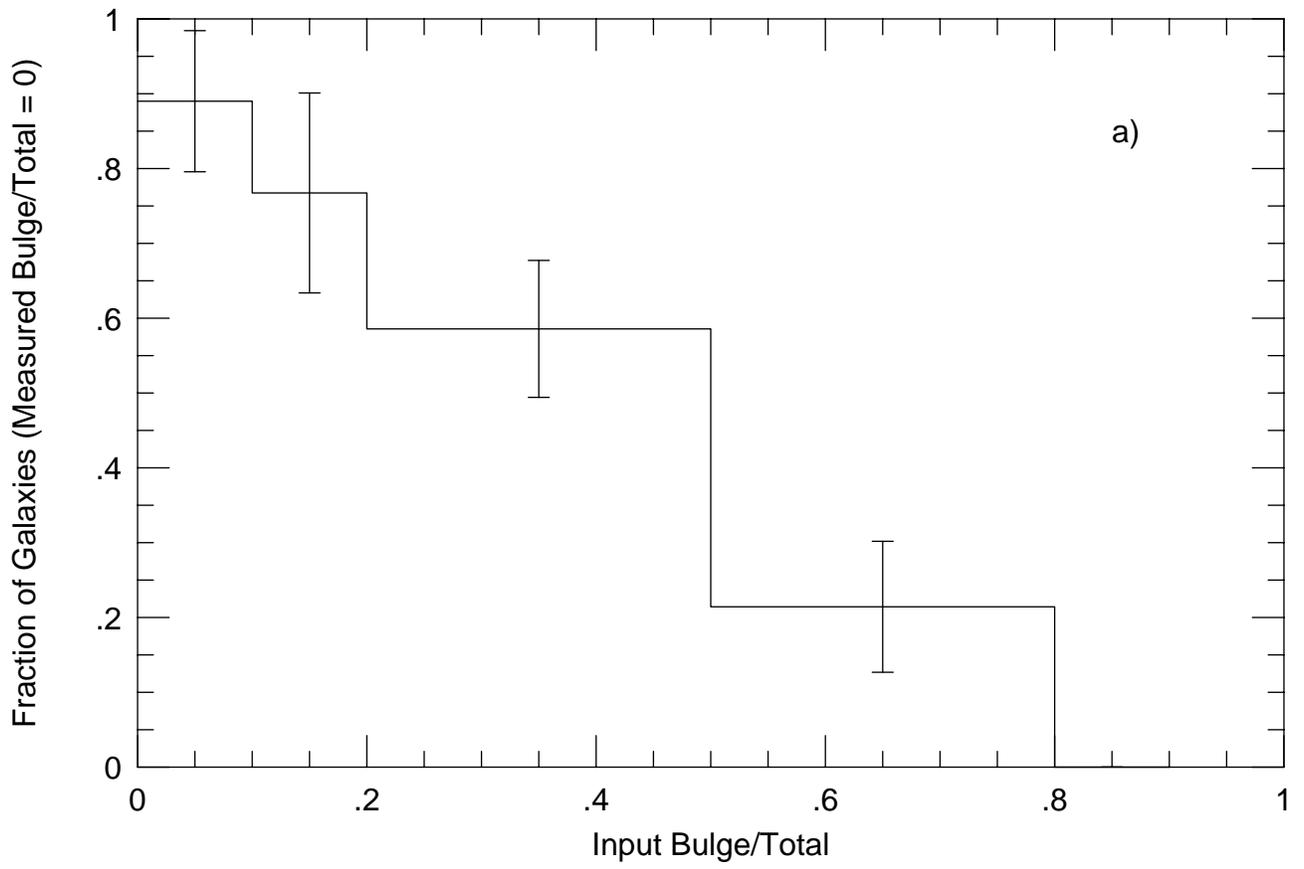
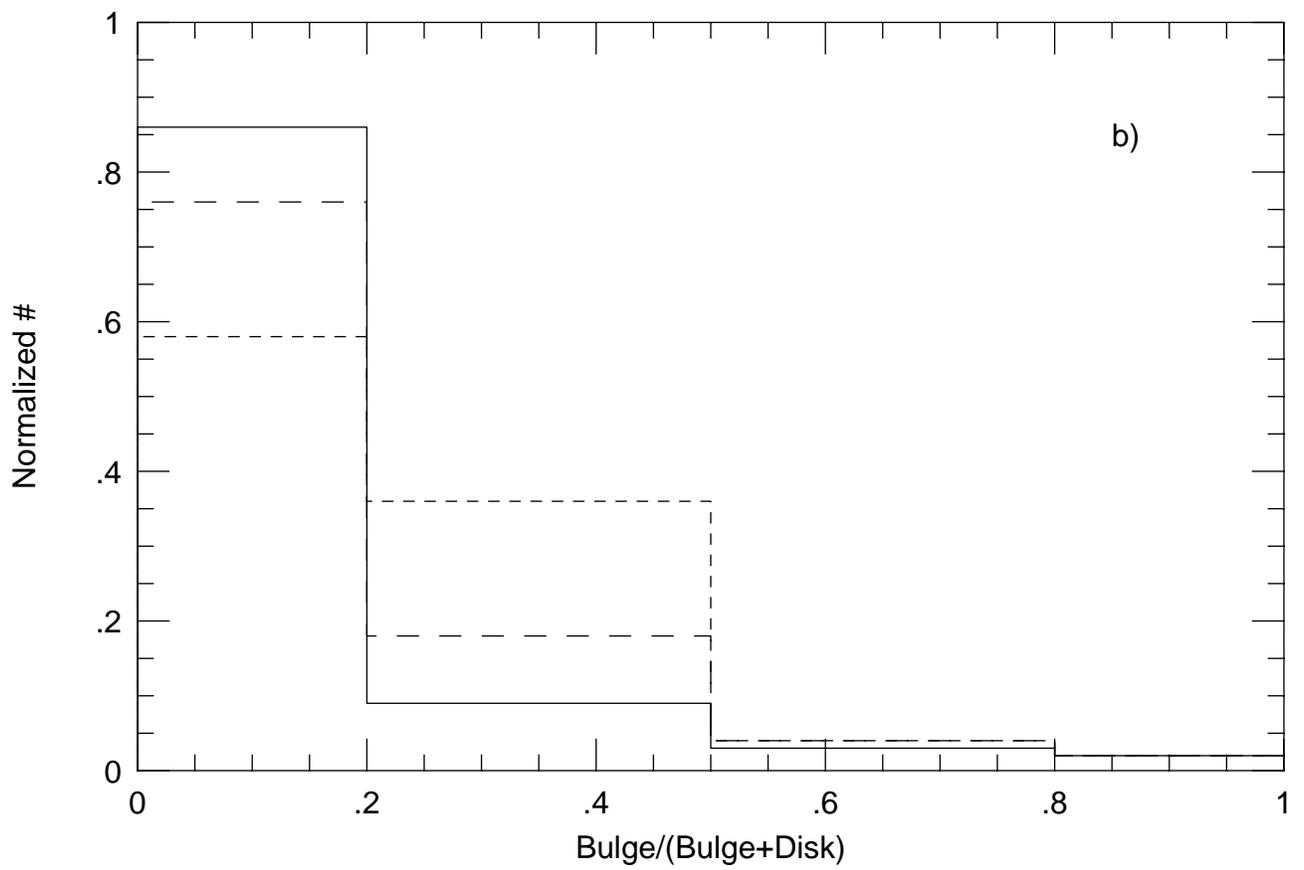
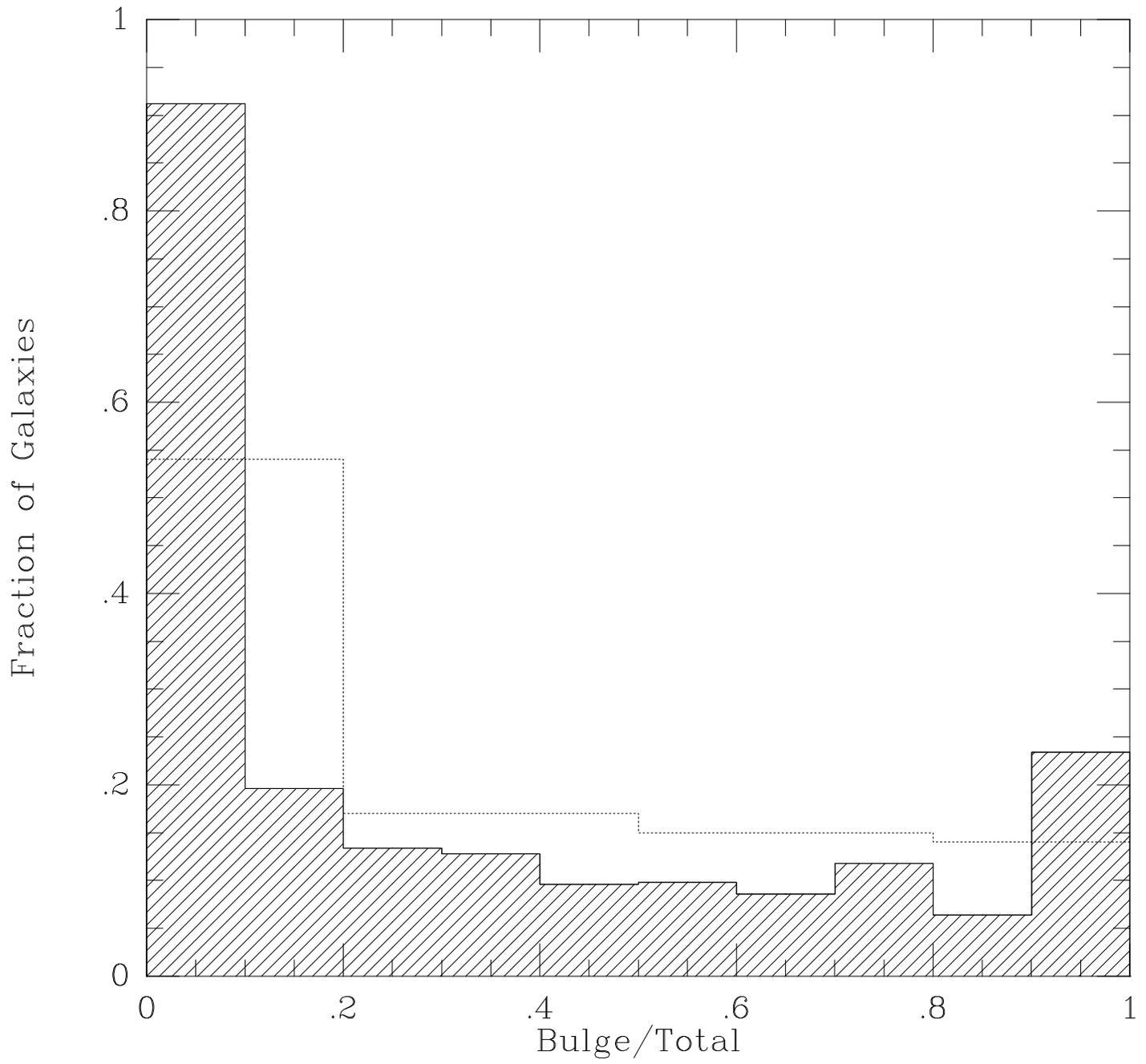
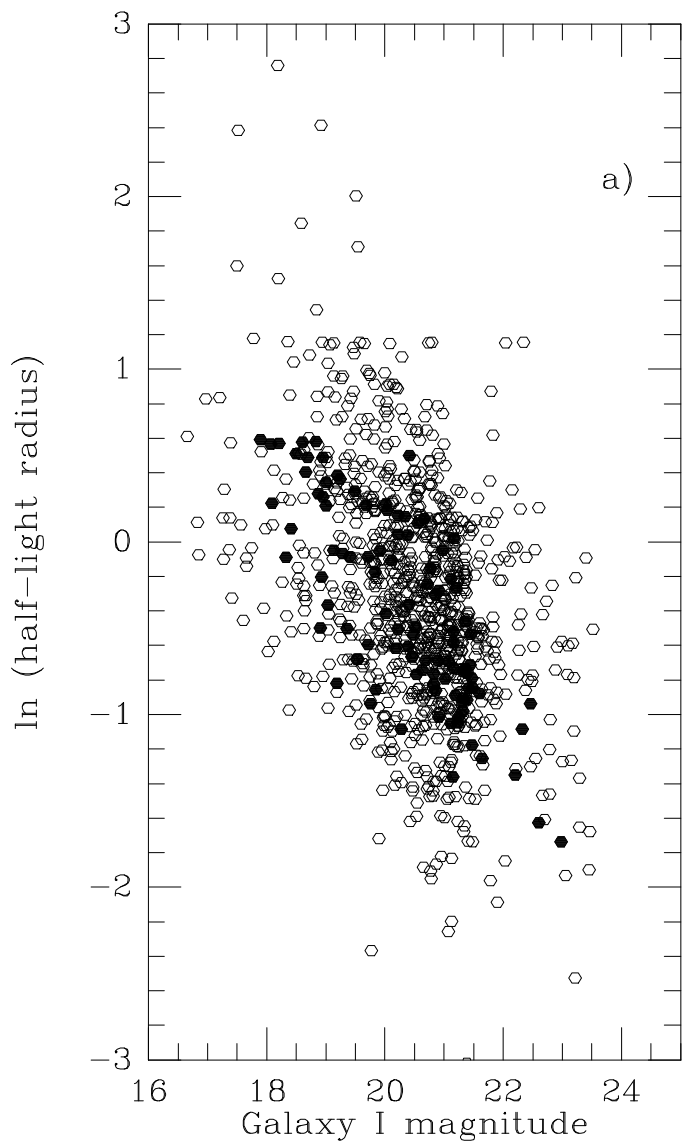
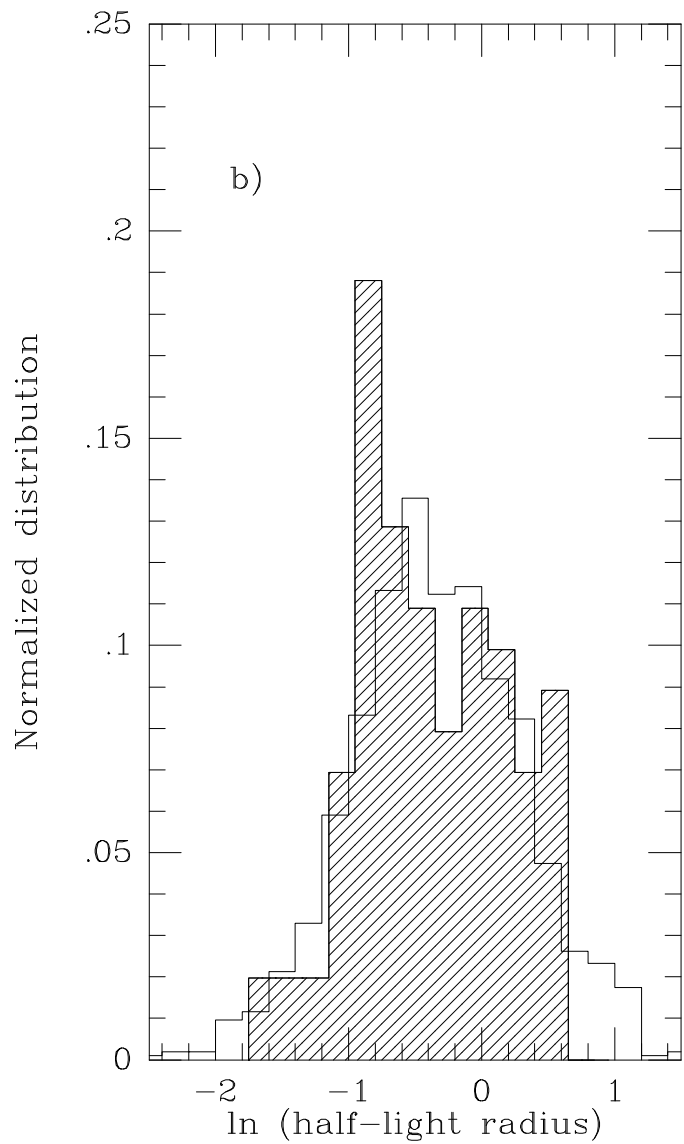
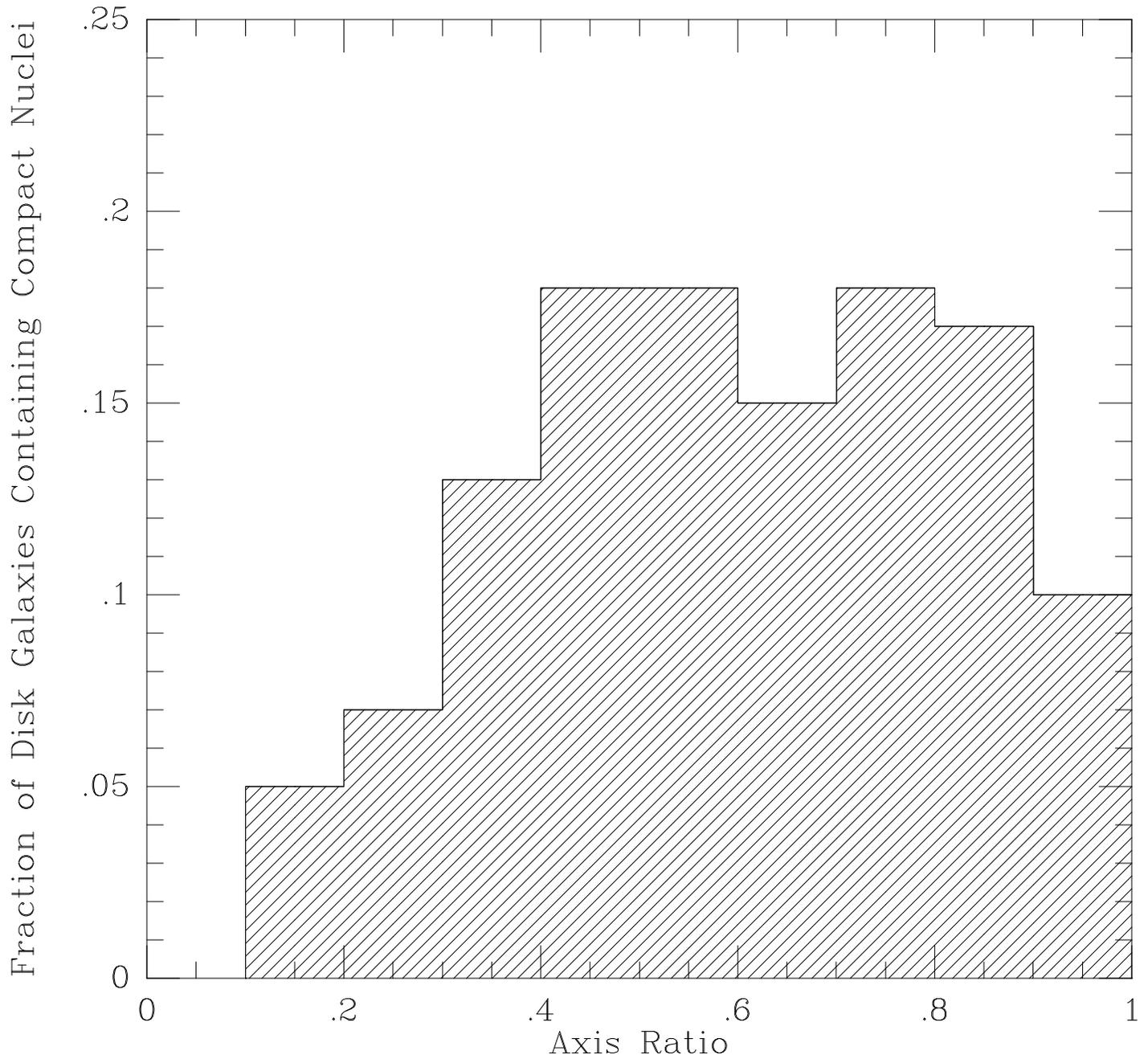
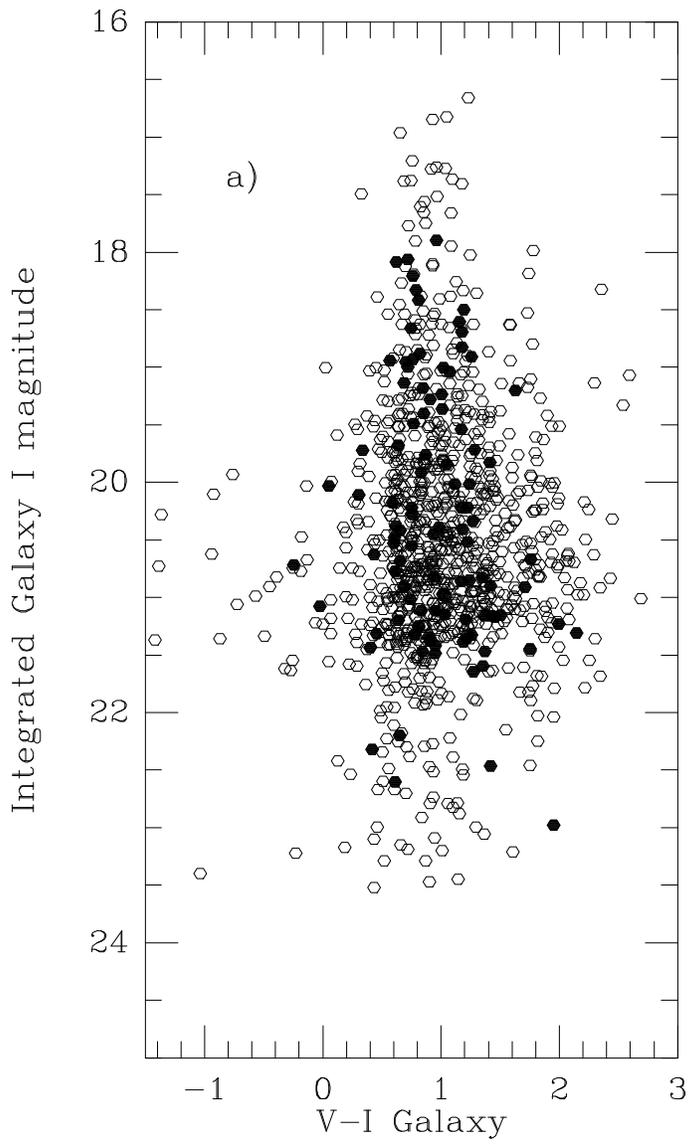
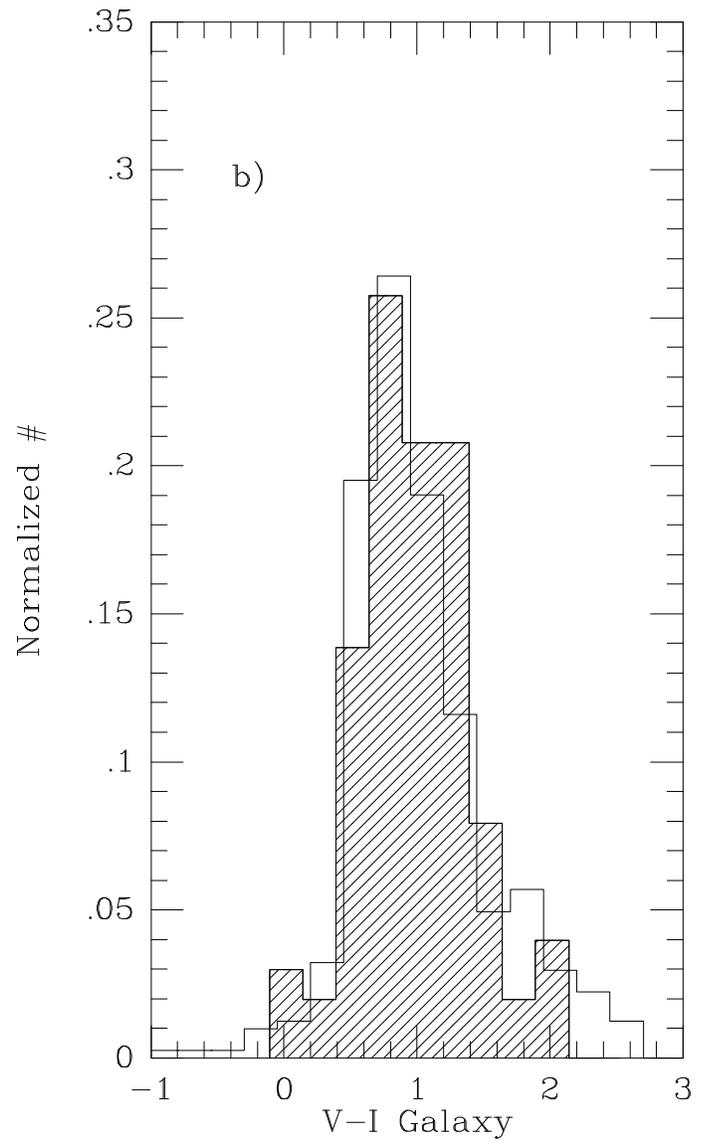
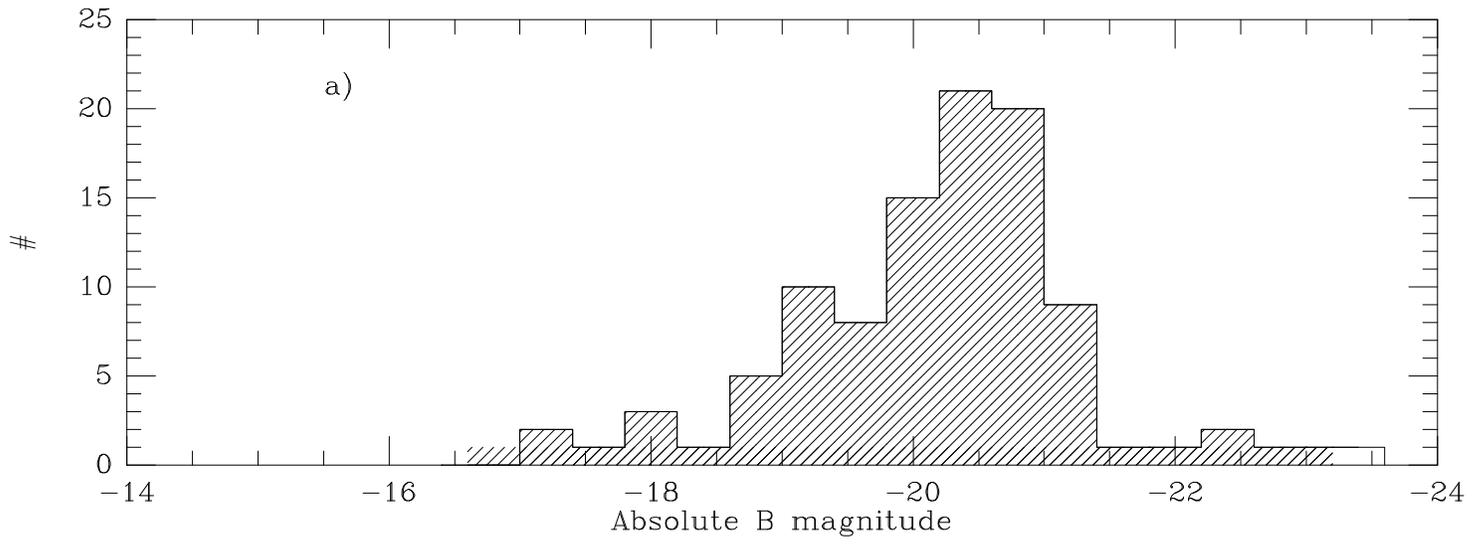
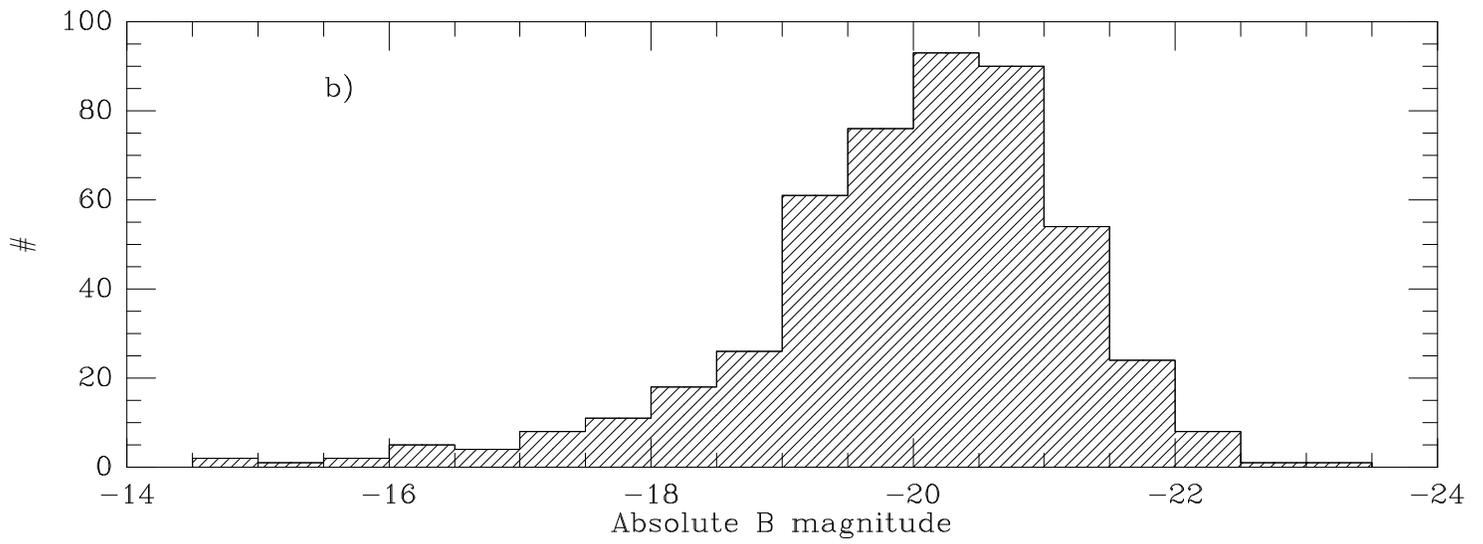
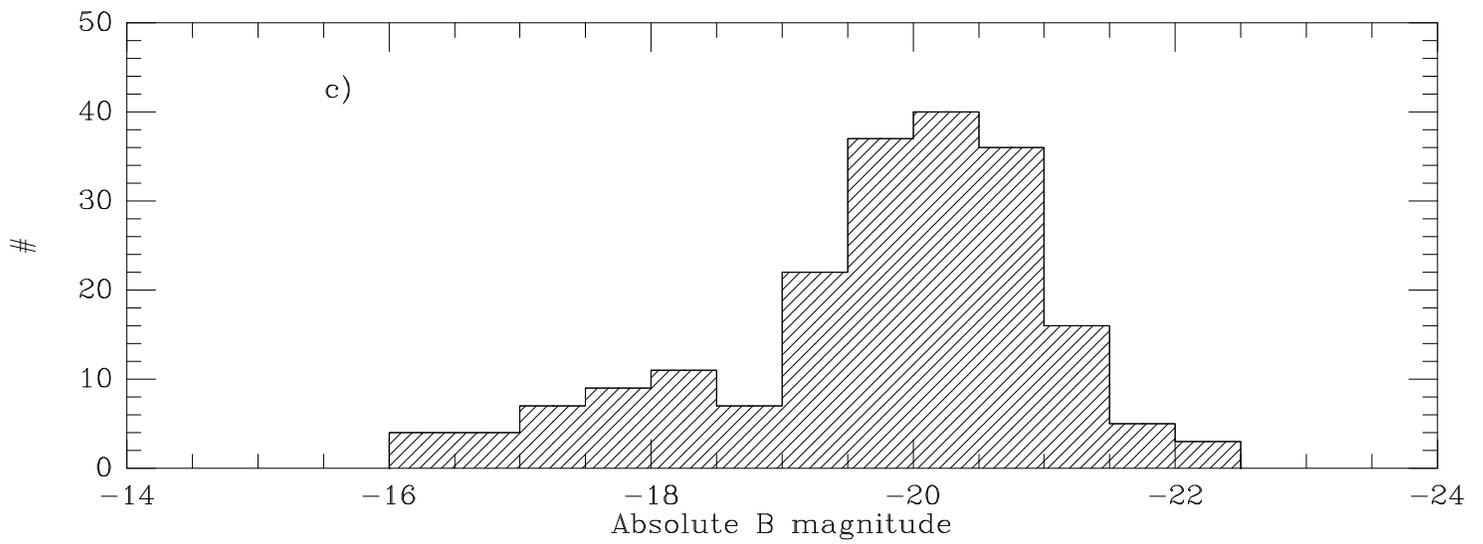
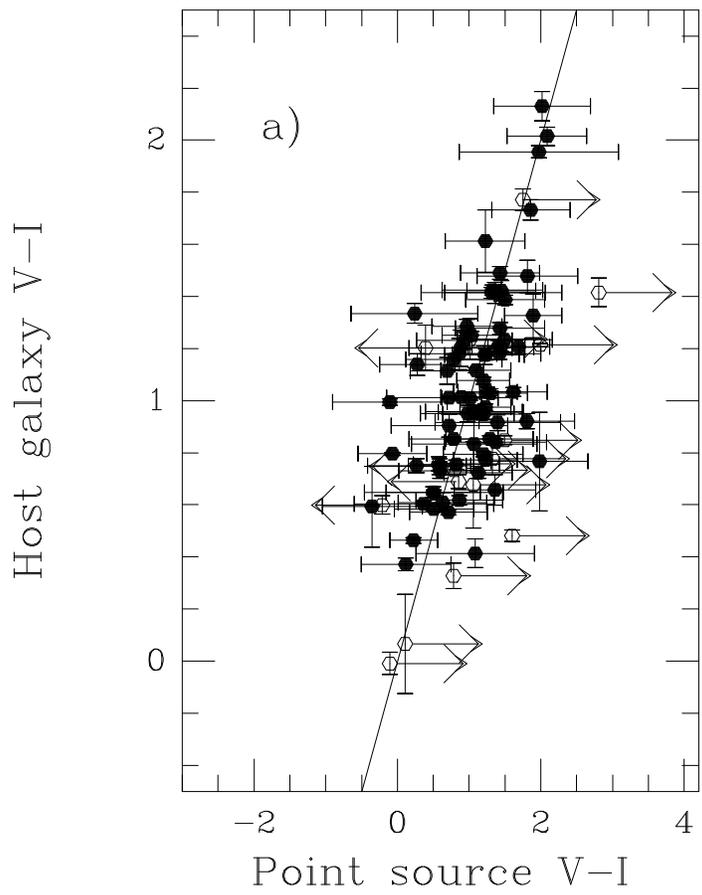
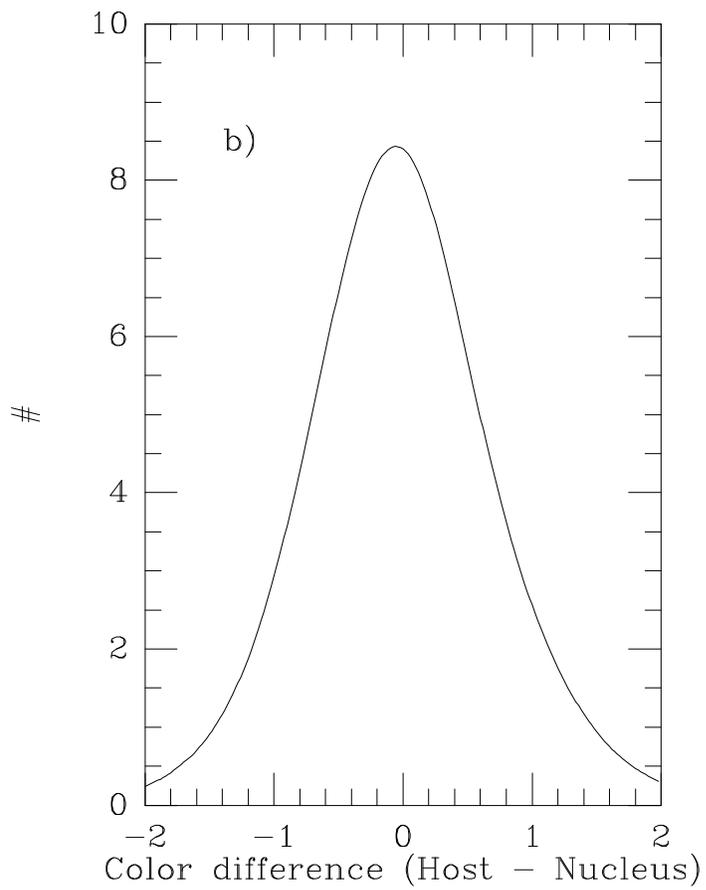
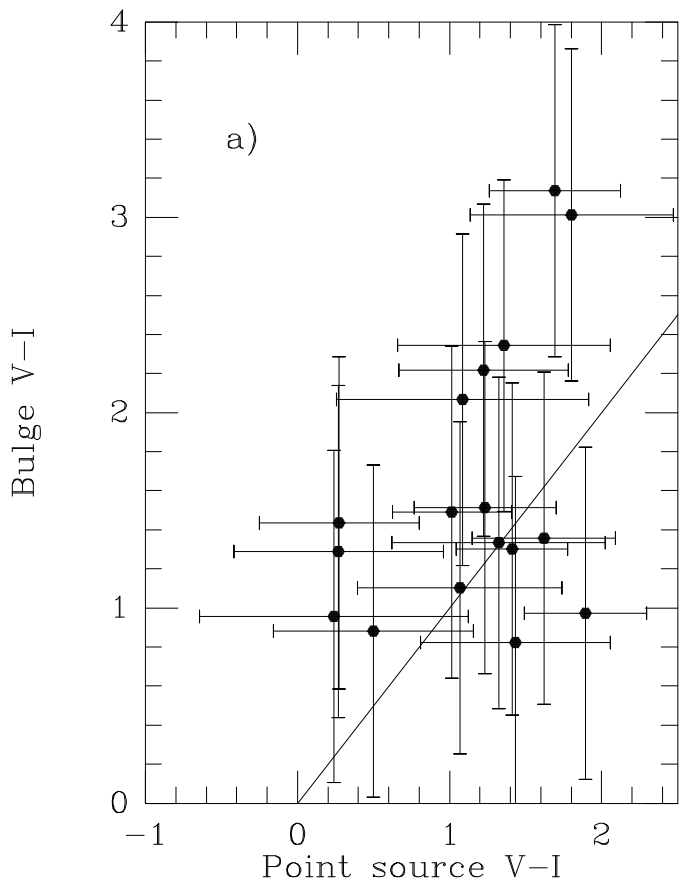
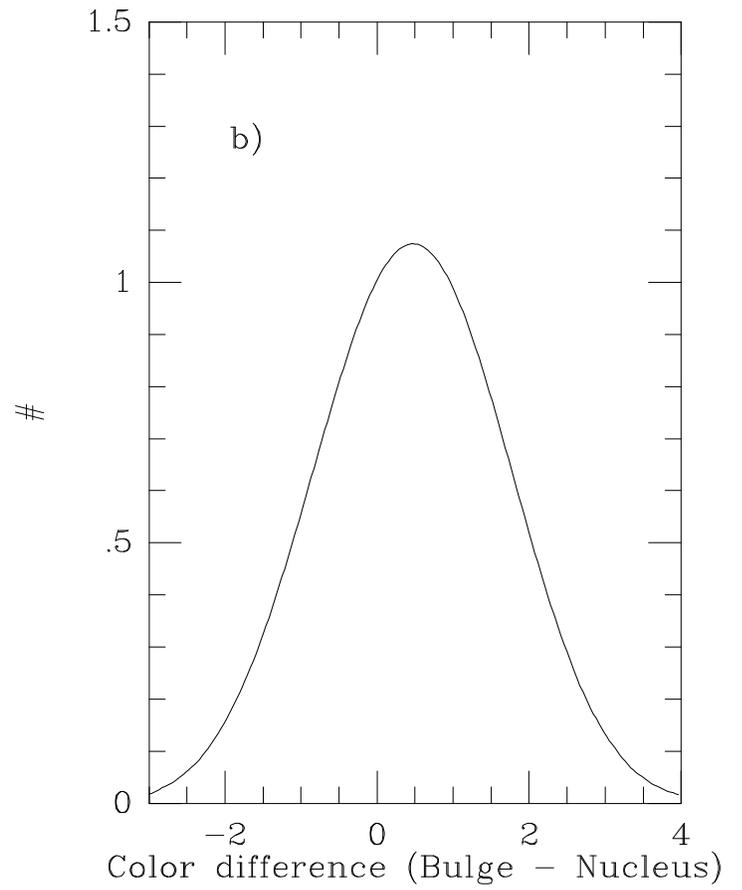
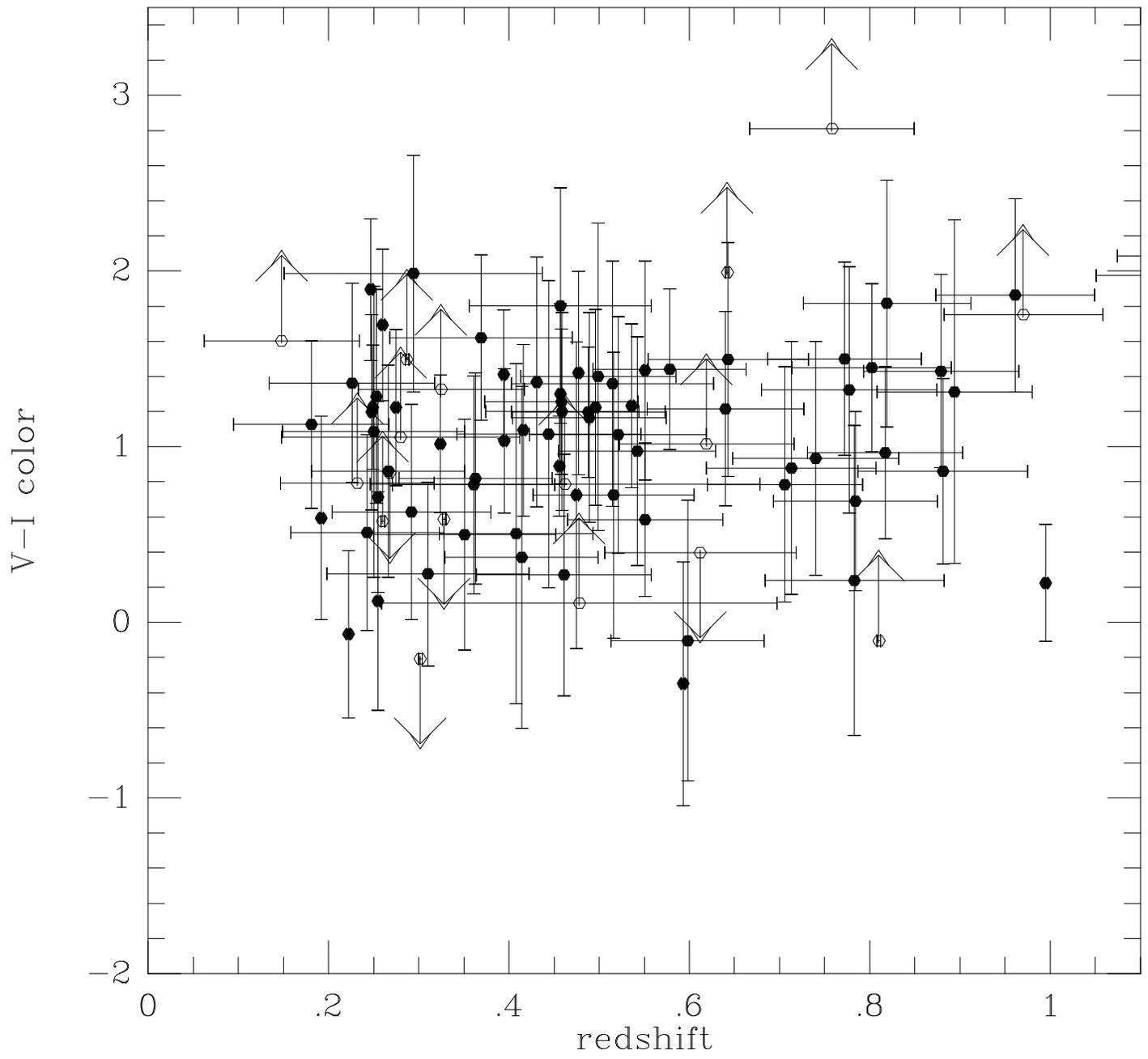
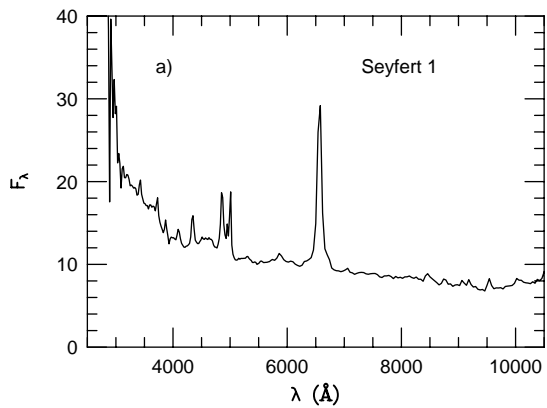
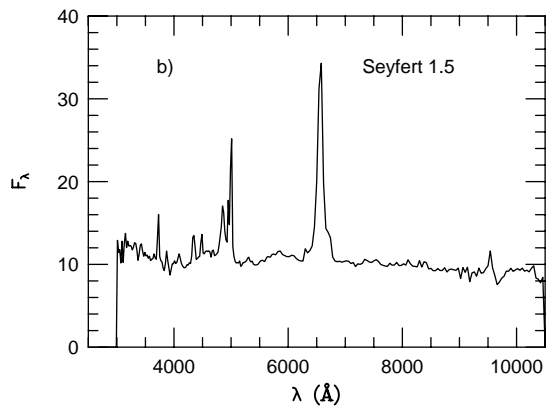
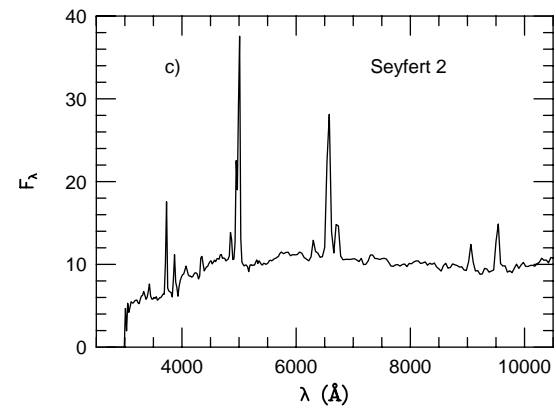
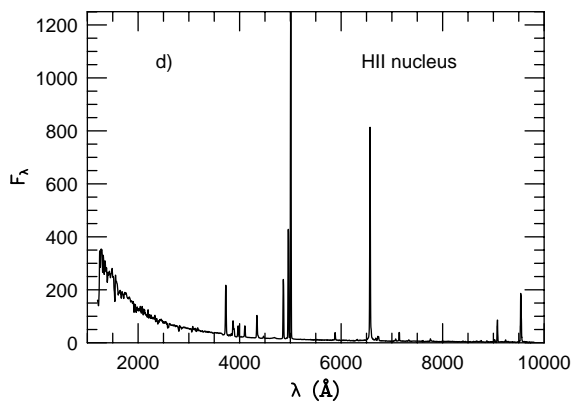
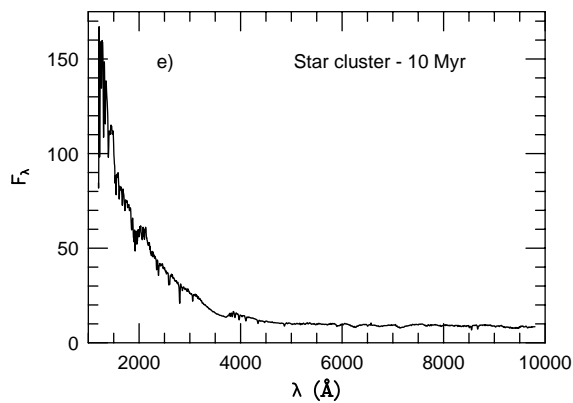
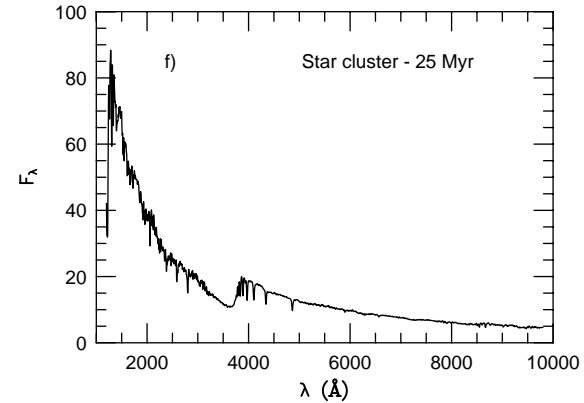
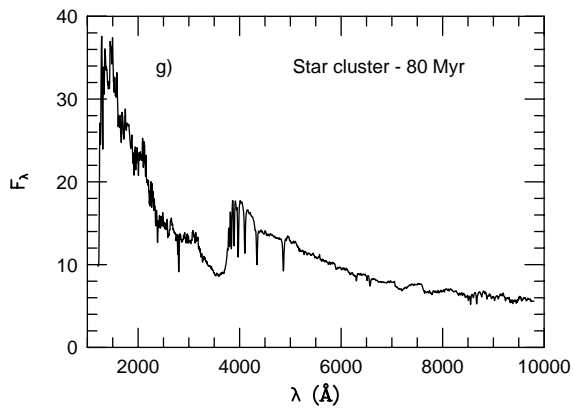
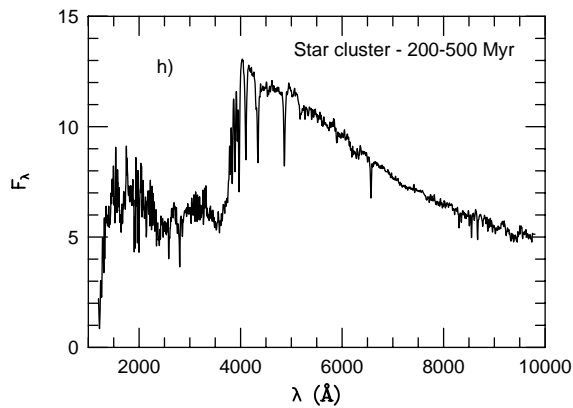
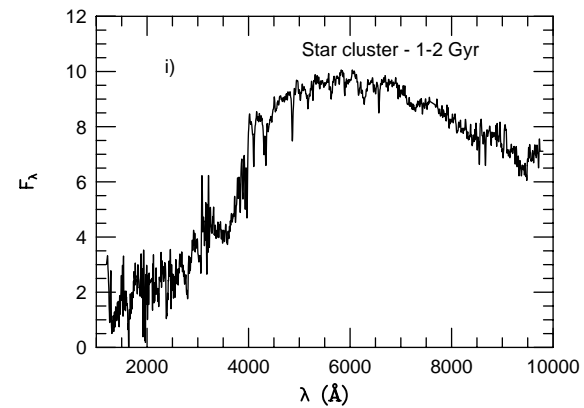
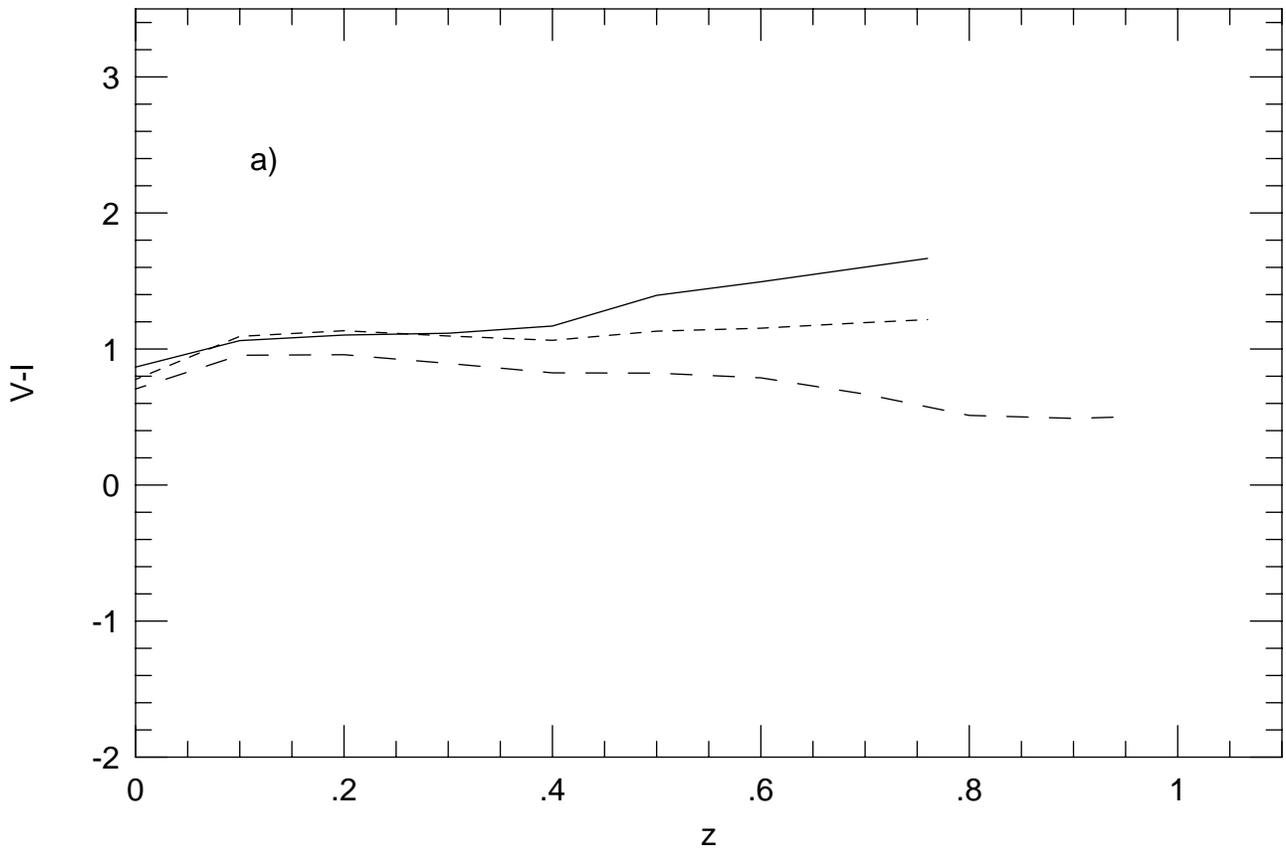
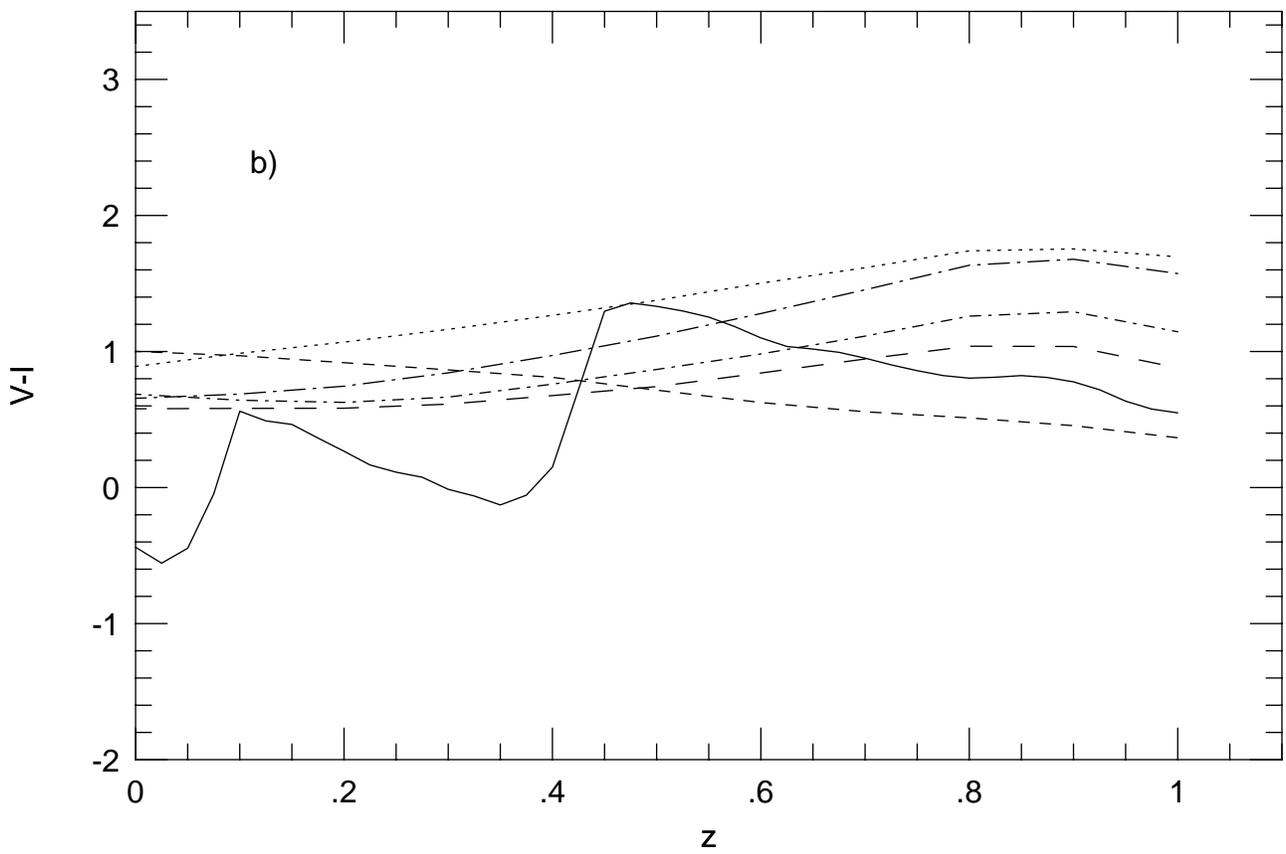
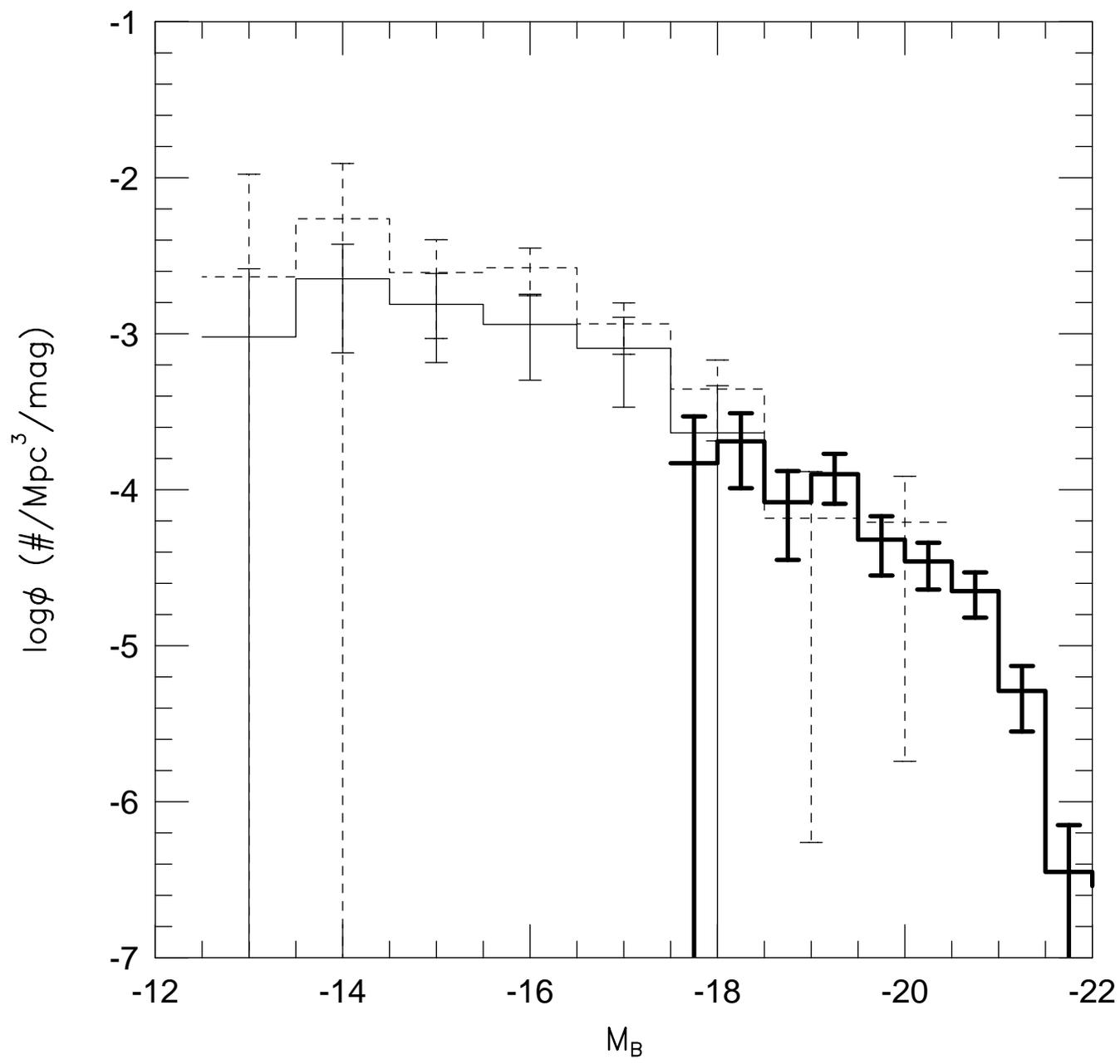
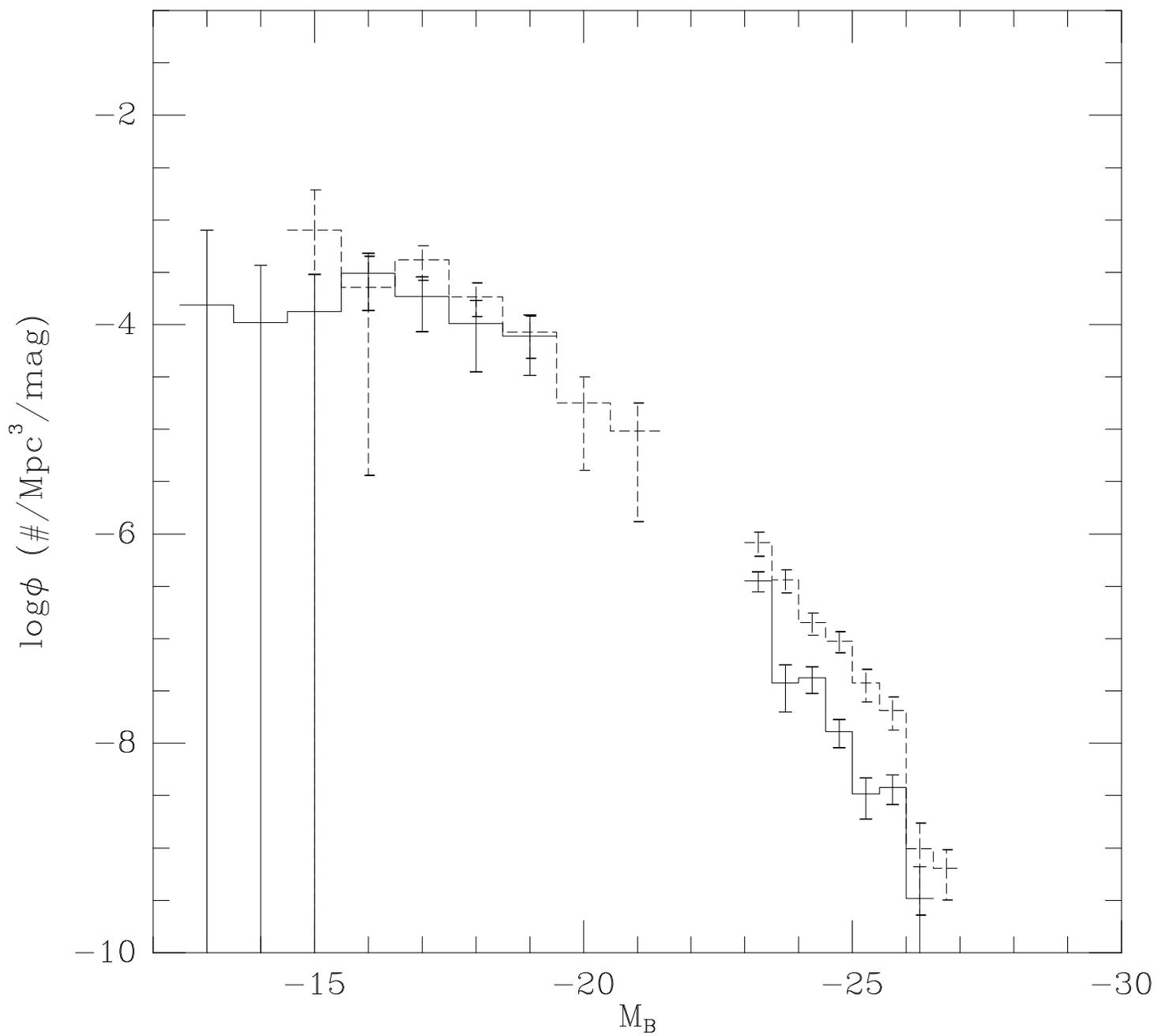
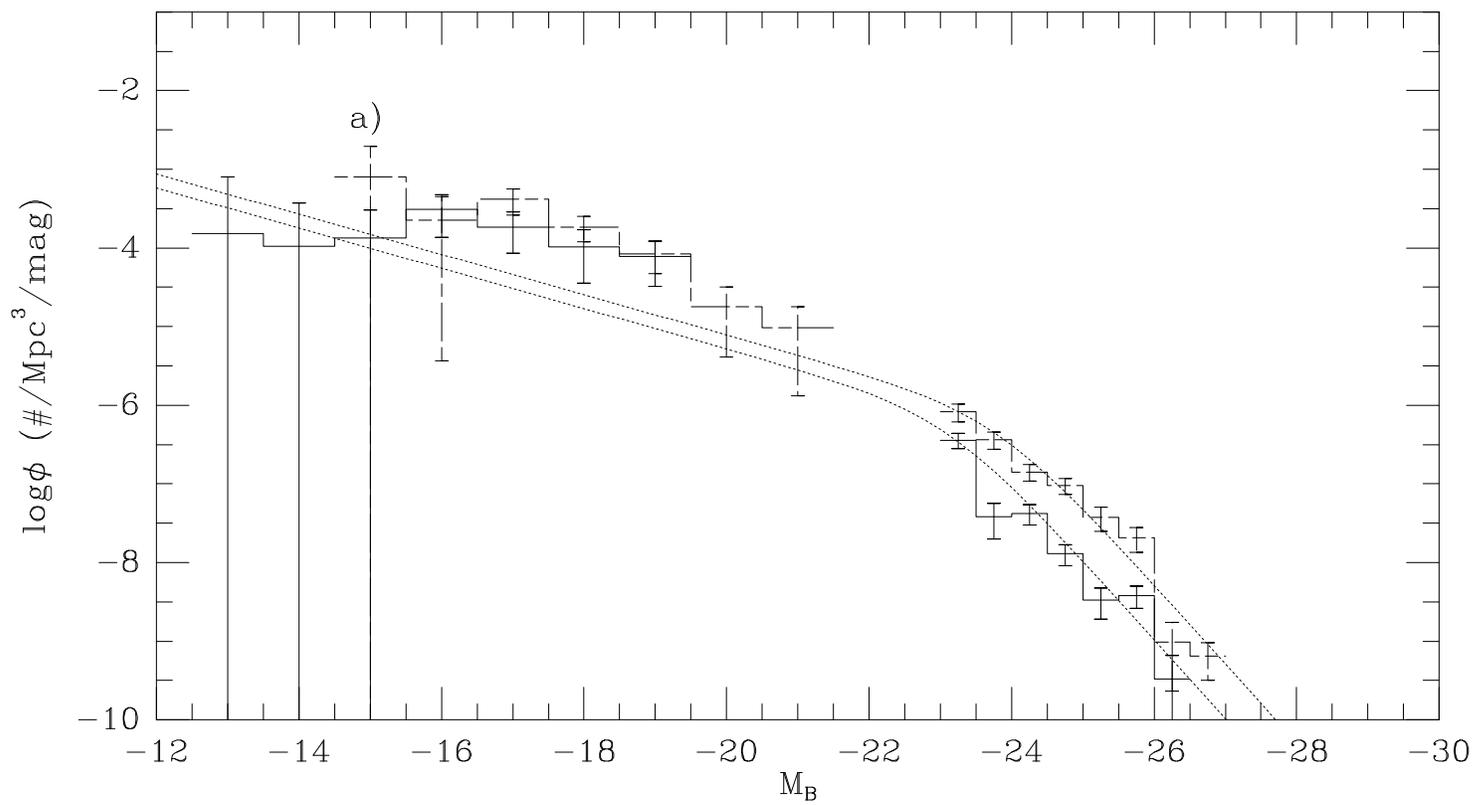
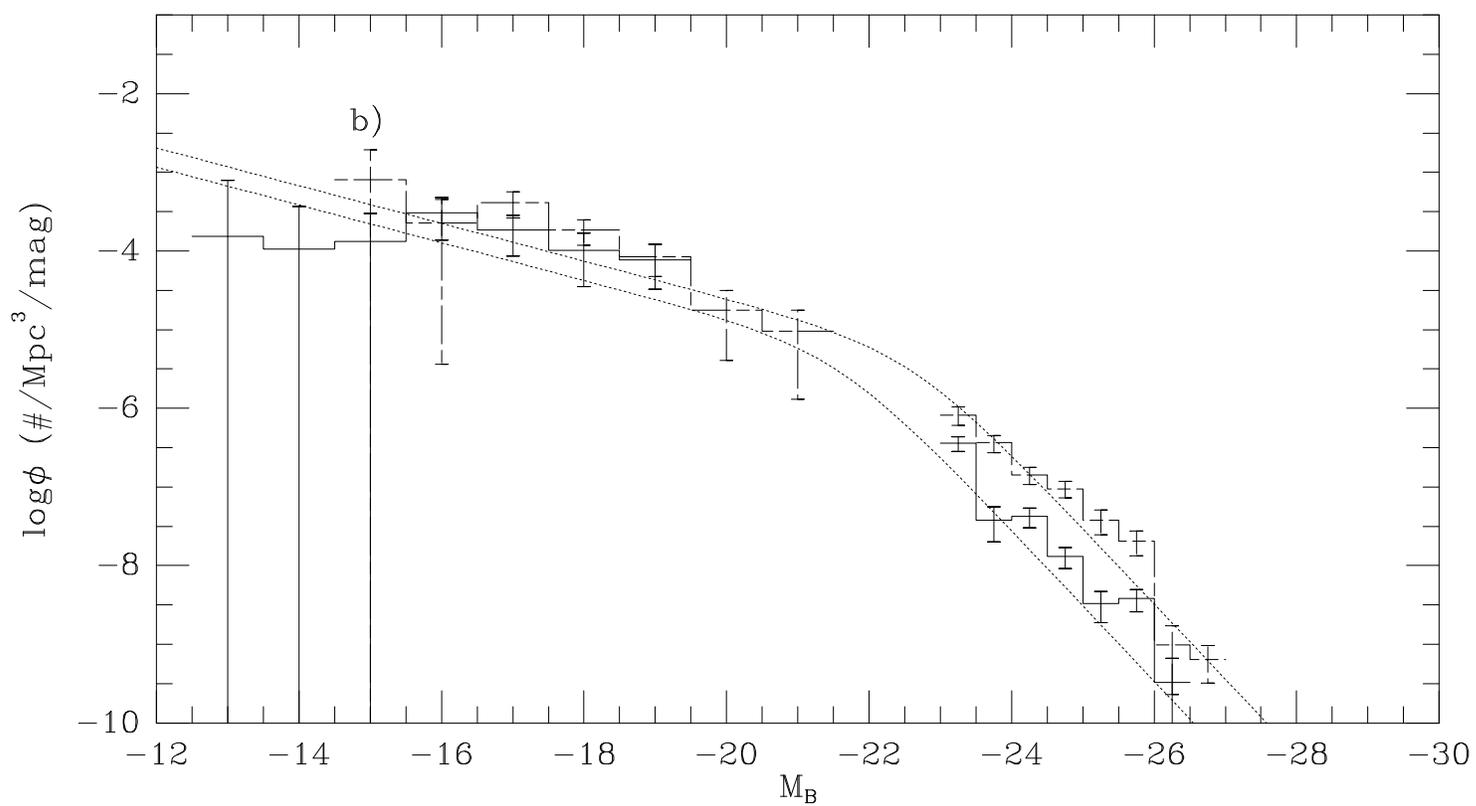
\end{document}